\documentclass[12pt]{JHEP3}

\usepackage{epsfig,amsfonts,amssymb,amsmath,cite}
\usepackage[all]{xy}

\newcommand{\bP}{\mathbb{P}}
\newcommand{\bZ}{\mathbb{Z}}

\newcommand{\cQ}{\mathcal{Q}}

\newcommand{\kk}{k}

\newcommand{\FF}{{\cal F}}

\newcommand{\NN}{{\cal N}}

\newcommand{\be}{\begin{equation}}
\newcommand{\ee}{\end{equation}}
\newcommand{\ben}{\begin{eqnarray}\displaystyle}
\newcommand{\een}{\end{eqnarray}}

\newcommand{\refb}[1]{(\ref{#1})}
\newcommand{\p}{\partial}

\newcommand{\non}{\nonumber}

\newcommand{\cN}{\mathcal{N}}
\newcommand{\cH}{\mathcal{H}}
\newcommand{\cM}{\mathcal{M}}
\def\Tr{\,{\rm Tr}\, }

\newcommand{\dx}{c}

\newcommand{\gref}{g_{\rm Coulomb}}

\newcommand{\QC}{Q_{\rm Coulomb}}
\newcommand{\bQC}{{\bar Q}_{\rm Coulomb}}

\newcommand{\OmS}{\Omega_{\rm S}} 
\newcommand{\bOmS}{\bar\Omega_{\rm S}}

\title{Generalized quiver mutations \\ and single-centered indices}
 
\preprint{ 
CERN-PH-TH/2013-221\\
HRI/ST/1304\\
arXiv:1309.7053v2}

\author{Jan Manschot$^{1}$, Boris Pioline$^{2,3}$, Ashoke Sen$^{4}$
\\
$^1$ {\it Institut Camille Jordan, Universit\'e Claude Bernard Lyon 1, \\ 43 boulevard du 11 novembre 1918, 69622 Villeurbanne cedex, France}\\

$^2$ {\it CERN PH-TH,
Case C01600, CERN, CH-1211 Geneva 23, Switzerland}\\

$^3$ {\it Laboratoire de Physique Th\'eorique et Hautes
Energies, CNRS UMR 7589, \\
Universit\'e Pierre et Marie Curie,
4 place Jussieu, 75252 Paris cedex 05, France} \\

$^4$ Harish-Chandra Research Institute,
Chhatnag Road, Jhusi, Allahabad 211019, India
\\

\vspace*{2mm} \email{
jan.manschot@univ-lyon1.fr, boris.pioline@cern.ch,
sen@hri.res.in}
\vspace*{-3mm}

} 

\abstract{Quiver quantum mechanics is invariant under Seiberg
  duality. A mathematical consequence is that the cohomology of the
  Higgs branch moduli space is invariant under mutations of the quiver. The Coulomb
  branch formula, on the other hand, conjecturally expresses  the
  Poincar\'e / Dolbeault polynomial of the Higgs branch moduli space in terms of
  certain quantities known as single-centered indices. In this work we
  determine the transformations of these single-centered 
indices under mutations. Moreover, we generalize these mutations to
quivers whose nodes carry single-centered indices different from unity. Although the Higgs branch
description of these generalized quivers is currently unknown, the Coulomb branch formula
is conjectured to be invariant under generalized mutations.}

\begin{document}

\tableofcontents

\section{Introduction and summary} \label{sintro}

Originally introduced in order to describe D-branes at orbifold singularities \cite{Douglas:1996sw},
quiver quantum mechanics has become a powerful tool for determining the spectrum of BPS states both in four-dimensional gauge theories with $\cN=2$ global 
supersymmetries
\cite{Fiol:2000pd,Fiol:2000wx,Alim:2011ae,Alim:2011kw,Cecotti:2012se,Xie:2012gd,Galakhov:2013oja,Cordova:2013bza,Chuang:2013wt,Cirafici:2013bha} and in four-dimensional type II string vacua with the same amount of
local supersymmetry \cite{Douglas:2000ah,Fiol:2000wx,Denef:2002ru,
Denef:2007vg,Aganagic:2010qr}. Physically, quiver quantum mechanics encodes the low energy dynamics of open strings stretched between D-brane constituents, and 
BPS bound states are identified as cohomology classes on the Higgs branch. Mathematically, the latter is interpreted as the moduli space of semi-stable quiver representations \cite{zbMATH00720513}. 

For quivers without oriented loops, such that the superpotential vanishes, the Higgs branch cohomology can be computed systematically \cite{1043.17010}. Equivalently, it can be computed on the Coulomb branch,  by studying the quantum mechanics of a set of point-like charged particles associated with the nodes of the quiver, and interacting by Coulomb and Lorentz-type forces according to the number of arrows between any two nodes \cite{Denef:2002ru}. The classical moduli space of such multi-centered solutions is a finite dimensional compact symplectic space \cite{deBoer:2008zn}, and the corresponding supersymmetric quantum mechanics \cite{Manschot:2011xc,Kim:2011sc,Lee:2011ph} can be solved using localization techniques \cite{Manschot:2010qz,Manschot:2011xc,Manschot:2013sya} (see \cite{Pioline:2013wta} for a recent review). Agreement between the two approaches for any choice of stability condition (equivalently, Fayet-Iliopoulos or FI
parameters) was demonstrated recently  
in \cite{Sen:2011aa,Manschot:2013sya}.

For quivers with loops, the situation is much more involved: on the Higgs branch side, there is currently no systematic way to compute the cohomology of a quiver with generic superpotential,
except for Abelian quivers which can be treated by ad hoc methods \cite{Bena:2012hf,Lee:2012sc,Lee:2012naa,Manschot:2012rx}. On the Coulomb branch side, the BPS phase space is in general no longer compact, due to the occurence of scaling solutions \cite{Denef:2007vg,Bena:2006kb} where three or more constituents approach each other at arbitrary small distance. While the symplectic volume of this phase space is still finite \cite{deBoer:2008zn,Manschot:2011xc},
the number of associated Coulomb branch states fails to match the number of states on the Higgs branch, by an exponential amount \cite{Denef:2007vg}. Based on the observation on simple cases that the discrepancy  originates solely from the middle cohomology  
(more precisely, the Lefschetz singlet part thereof) and is insensitive to wall-crossing \cite{Bena:2012hf}, it was proposed in \cite{Manschot:2012rx} that the
isomorphism between the Coulomb and Higgs branch could be restored by
postulating the existence of new Coulomb branch constituents, behaving
as  point-like particles 
carrying composite charge $\gamma$ and internal degrees of freedom  with index
$\OmS(\gamma)$, insensitive  to the choice of stability condition. 
Conjecturally, the Poincar\'e-Laurent polynomial 
of the quiver moduli space (defined in \eqref{epol} below) 
is expressed
in terms of these invariants, known as single-centered indices (or indices
associated with pure Higgs, or intrinsic Higgs states)
through the Coulomb branch formula (see \eqref{essp1}). 
Defining and computing the single-centered indices  $\OmS(\gamma)$ directly
remains an open problem.

While there is no general prescription for computing the
Poincar\'e-Laurent polynomial of 
a quiver with generic superpotential, it  is known to be invariant
under specific transformations of the quiver known as mutations
\cite{zbMATH05573998,ks,zbMATH05848698}. Quiver mutation was
first introduced in the context of ADE quivers \cite{Bernstein:1973},
and is one of the basic principles of the theory of cluster algebras \cite{2001math4151F}.  
In terms of the quiver quantum mechanics descriptions of BPS bound
states, mutations  are a manifestation of  Seiberg  
duality \cite{Feng:2001bn,Beasley:2001zp,Berenstein:2002fi,Feng:2002kk,Mukhopadhyay:2003ky,Herzog:2004qw,Vitoria:2007ff}, and arise when the splitting between BPS and anti-BPS
states is varied \cite{Denef:2000nb,Aganagic:2010qr,Andriyash:2010yf,Cordova:2013bza}.
This happens in particular when the moduli are varied around a point where 
one of the constituents of the bound state becomes massless, and is responsible
for the monodromy transformation of the BPS spectrum \cite{Aganagic:2010qr,Andriyash:2010yf}. A natural question is  to determine the action of mutations on the
single-centered invariants  $\OmS(\gamma)$ appearing in the Coulomb branch formula. 

From the point of view of the Coulomb branch formula,  however, quiver
moduli spaces are but a very special case where the basis vectors
associated to the nodes of the quiver carry unit 
index, $\OmS(\gamma_i)=1$ and $\OmS(\ell\gamma_i)=0$ if $\ell>1$ (mathematically, the nodes represent spherical objects in the derived category of representations). Formally, one could very well keep the same quiver topology but associate different indices $\OmS(\gamma_i)$ to the
basis vectors and multiples thereof, and use the Coulomb branch formula to produce a set
of  symmetric 
Laurent polynomials satisfying the standard wall-crossing
properties. We refer to such quivers with non-standard single-centered indices 
as generalized quivers, and to the corresponding Laurent polynomials as 
generalized quiver invariants. 
Ref. \cite{Manschot:2011xc} showed that, in the case of quivers without closed loops,
such generalized quivers appear in wall-crossing 
formulas for Donaldson-Thomas invariants \cite{ks,MR2951762}. 
Whether or not the generalized quiver invariants correspond to
the Poincar\'e/Dolbeault polynomial of a putative moduli space is unclear
to us at this stage, but we can ask whether invariance under mutations
can be extended to this set of polynomials. A suggestive
fact is that mutations can also be defined for
cluster algebras with skew-symmetrizable -- as opposed to skew-symmetric -- exchange matrix,
which are naturally represented by quivers with multiplicity \cite{zbMATH05145743,
2003math11245F,Labardini:2013}. 

Another reason to expect such a generalization is the physical `Fermi flip' picture of mutation 
developed in the context of split attrator flows in supergravity in \cite{Andriyash:2010yf}. Namely,
in the vicinity of certain walls in moduli space (conjugation walls in the language of \cite{Andriyash:2010yf}, 
or walls of the second kind in the language of \cite{ks}), the representation of a BPS state of total charge $\gamma=\gamma_j+N \gamma_k$ as a halo of particles  carrying charges $\ell_i\gamma_k$ with $\ell_i>0$
orbiting around a core of charge $\gamma_j$ can become invalid, and needs to be replaced by a halo of particles carrying charges $-\ell_i\gamma_k$ with $\ell_i>0$ 
around a core of
charge $\gamma_j+M_j \gamma_k$, for some positive integer $M_j$ 
\cite{Andriyash:2010yf}.  
This is possible when the particles of charge $\ell\gamma_k$
behave as fermions (i.e. carry positive\footnote{Due to the supermultiplet structure 
a state with positive index behaves as a fermion while forming a bound state \cite{Denef:2002ru}.} 
index), so that the Fermi vacuum can be replaced
by the filled Fermi sea. 
In this paper, we shall argue 
that this picture applies just as well for generalized quivers with oriented loops, and naturally
suggests that the Laurent polynomials produced by the Coulomb branch formula are invariant
under a generalized mutation transformation. Before stating this transformation,
we need to set up some notations.

\subsection{Review of quiver invariants and the Coulomb branch formula}

Consider a quiver with $K$ nodes with dimension vectors $(N_1,\cdots N_K)$, 
stability (or Fayet-Iliopoulos, or FI) parameters $(\zeta_1,\cdots \zeta_K)$ satisfying $\sum_{i=1}^K N_i \zeta_i=0$, and $\gamma_{ij}$ arrows from the $i$-th node to the $j$-th node. 
We denote such a quiver by $\cQ(\gamma;\zeta)$, where $\gamma$
is a vector  $\gamma=\sum_{i=1}^K
N_i\gamma_i$ in a $K$-dimensional lattice $\Gamma$ spanned by basis vectors $\gamma_i$ 
associated to each node. We shall denote by $\Gamma^+$ the collection of lattice vectors
of the form $\sum_i n_i\gamma_i$ with $n_i\ge 0$; clearly all physical quivers are 
described by some vector $\gamma\in\Gamma^+$. 
We introduce a bilinear symplectic product (the Dirac-Schwinger-Zwanziger, or DSZ product)
on $\Gamma$ via
$\langle \gamma_i, \gamma_j\rangle=\gamma_{ij}$.
To define the quiver moduli space, we introduce   complex
variables $\phi_{\ell\kk, \alpha, ss'}$ for every pair $\ell,\kk$ for which 
$\gamma_{\ell\kk}>0$. Here $\alpha$ runs over $\gamma_{\ell\kk}$ values,
$s$ is an index labelling the
fundamental representation of $U(N_\ell)$ and $s'$ is an index representing
the anti-fundamental representation of $U(N_{\kk})$. The moduli space $\cM(\gamma;\zeta)$ of
classical vacua is the  space of solutions to  the  D-term and F-term constraints,
\be \label{emodi1}
\begin{split}
 \sum_{\kk , s,t,s'\atop \gamma_{\ell\kk}>0} \phi_{\ell\kk, \alpha, ss'}^* \, T^a_{st} \, 
\phi_{\ell\kk,\alpha,t s'} - \sum_{\kk ,s,t,s'\atop \gamma_{\kk\ell}>0} 
\phi_{\kk\ell, \alpha, s's}^* 
\, T^a_{st} \, 
\phi_{\kk\ell,\alpha,s't} = \zeta_\ell \, \Tr(T^a)\quad \forall \, \ell, \, a \, , & \\
 {\p W\over \p \phi_{\ell\kk,\alpha,ss'}}=0\, \ , &
\end{split}
\ee
modded out by the natural action of the gauge group $\prod_\ell U(N_\ell)$.
Here $T^a$'s are the generators of the $U(N_\ell)$ gauge group, and $W$ is a
generic gauge invariant superpotential holomorphic in the variables
$ \phi_{\ell\kk, \alpha, ss'}$. For generic potential, $\cM(\gamma;\zeta)$ is a compact
algebraic variety, which is smooth if the vector $\gamma$ is primitive.

Let $Q(\gamma; \zeta;y)$ be the Poincar\'e-Laurent polynomial
of the quiver moduli space $\cM(\gamma;\zeta)$,
\be \label{epol}  
Q(\gamma; \zeta;y)  = \sum_{p=0}^{2d} b_p(\cM)\, (-y)^{p-d}
\ee
where $d$ is the complex 
dimension of $\cM$
and the $b_p(\cM)$'s are the topological Betti numbers of $\cM$.
The Coulomb branch formula 
for $Q(\gamma; \zeta;y)$,  
which we denote by $\QC(\gamma; \zeta;y)$, takes the form \cite{Manschot:2011xc,
Manschot:2012rx,Manschot:2013sya}
\be
 \label{essp1}
\begin{split}
\QC(\gamma; \zeta;y) =& \sum_{m|\gamma} 
\frac{\mu(m)}{ m}  {y - y^{-1}\over y^m - y^{-m}}
\bQC(\gamma/m; \zeta;y^m) \ , \\
\bQC(\gamma; \zeta;y) =& 
\sum_{n\ge 1}\sum_{\{\alpha_i\in \Gamma^+\} \atop \sum_{i=1}^n \alpha_i =\gamma}
 \frac{\gref\left(\{\alpha_1, \cdots, \alpha_n\},
 \{\dx_1,\cdots \dx_n\};y\right)}
{  |{\rm Aut}(\{\alpha_1,\cdots, \alpha_n\})|}
\\ &\quad
\prod_{i=1}^n \left\{\sum_{m_i\in\bZ\atop m_i|\alpha_i}
{1\over m_i} {y - y^{-1}\over y^{m_i} - y^{-m_i}}\, 
\Omega_{\rm tot}(\alpha_i/m_i;y^{m_i})
\right\}
\, ,
\end{split}
\ee
where $\mu(m)$ is the M\"obius function, 
$|{\rm Aut}(\{\alpha_1,\cdots \alpha_n\})|$ is a symmetry factor given by
$\prod_k s_k!$ if among the set $\{\alpha_i\}$ there are
$s_1$ identical vectors $\tilde \alpha_1$, $s_2$ identical vectors
$\tilde\alpha_2$ etc., and $m|\alpha$ means that $m$ is a common divisor of
$(n_1,\cdots , n_K)$ if $\alpha =\sum_\ell n_\ell \gamma_\ell$.
The sums over $n$ and $\{\alpha_1,\cdots \alpha_n\}$ in the second
equation label all possible ways of
expressing $\gamma$ as (unordered) sums of elements $\alpha_i$ of $\Gamma^+$. 
The coefficients
$\dx_i$ are determined in terms of the FI parameters $\zeta_i$ by $\dx_i
=\sum_\ell A_{i\ell} \zeta_\ell$ whenever  $\alpha_i=\sum_\ell A_{i\ell}\gamma_\ell$. 
From the restrictions $\sum_i \alpha_i
=\gamma$ and $\sum_\ell N_\ell \zeta_\ell=0$ it follows that 
$\sum_i \dx_i=0$.
The functions $\gref(\{\alpha_1,\cdots, \alpha_n\};
 \{\dx_1,\cdots \dx_n\};y)$, known as Coulomb indices, 
can be computed from the sum over
collinear solutions to Denef's equations for multi-centered black hole 
solutions \cite{Manschot:2011xc}. The functions $\Omega_{\rm tot}(\alpha;y)$ 
are expressed
in terms of the single-centered BPS invariants $\OmS$ through
\be \label{essp2}
\Omega_{\rm tot}(\alpha;y) = \OmS(\alpha;y) + 
\sum_{\{\beta_i\in \Gamma^+\}, \{m_i\in\bZ\}\atop
m_i\ge 1, \, \sum_i m_i\beta_i =\alpha}
H(\{\beta_i\}; \{m_i\};y) \, \prod_i 
\OmS(\beta_i;y^{m_i})
\, .
\ee
The $H(\{\beta_i\}; \{m_i\};y)$ are determined recursively using the
minimal modification hypothesis described in \cite{Manschot:2012rx}, 
and $\OmS(\alpha;y)$ are
expected to be $y$-independent constants for quivers with generic superpotential. 
A fully explicit recursive algorithm for computing the Coulomb indices  $\gref$
and $H$-factors was given in \cite{Manschot:2013sya}. 

In \cite{Manschot:2012rx} we also proposed a formula for the Dolbeault polynomial
\be
Q(\gamma; \zeta; y;t) \equiv \sum_{p,q} h^{p,q}(\cM) \, (-y)^{p+q-d} \, t^{p-q}\, ,
\ee 
where $h^{p,q}(\cM)$ are the Hodge numbers of $\cM$. The formula takes the same
form as \eqref{essp1}, \eqref{essp2}, with the only difference that $\OmS$ is
allowed to depend on $t$, and 
the arguments $y$ and $y^m$ inside $\QC$, $\bQC$, $\Omega_{\rm tot}$
and $\OmS$ are replaced by $y;t$ and $y^m;t^m$ respectively.\footnote{Eventually
we drop the $y$-dependence of $\OmS$ for quivers with generic superpotential.} 
The Coulomb indices $\gref$ and the functions $H$ remain
unchanged and 
independent of $t$.

\subsection{Generalized quivers and generalized mutations} \label{sgenmut}

We are now ready to state our main result. 
As mentioned above,
the Coulomb branch formula given  
in eqs.\eqref{essp1}, \eqref{essp2} leads to a set of symmetric Laurent polynomials satisfying  the standard wall-crossing formula, for any choice of symmetric Laurent polynomials 
$\OmS(\gamma;y;t)$. For ordinary quivers with a generic superpotential, the single-centered
invariants satisfy 
\be
\label{OmSquivers}
\OmS(n_i\gamma_i+n_j \gamma_j;y;t)=\begin{cases} 1 & \mbox{if}\   n_i=1, n_j=0\\
1 & \mbox{if}\  n_i=0, n_j=1\\
0 & \mbox{otherwise}
\end{cases}
\ee
for any  linear combination of two  basis vectors $n_i\gamma_i+n_j \gamma_j$. 
We refer to quivers equipped with more general choices of the single-centered invariants $\OmS(\gamma;y;t)$,
subject to the condition that they vanish unless $\gamma\in\Gamma^+$, 
 as `generalized quivers'.  

For such a generalized quiver, we introduce a 
generalized mutation $\mu_k^{\varepsilon}$ (where $\varepsilon=1$ for a right mutation, 
and $\varepsilon=-1$ for a left mutation) with respect to the $k$-th node, through
the following  transformation
rules of the basis vectors $\gamma_i$,
DSZ matrix $\gamma_{ij}$,  stability parameters $\zeta_i$, 
and dimension vector $N_i$:
\be
\label{mutDSZgen0}
\begin{split}
\gamma'_i=&\begin{cases}
-\gamma_k & \hbox{if $i=k$} \\
\gamma_i + M \, {\rm max}(0,\varepsilon \gamma_{ik})\, \gamma_k&  \hbox{if $i\neq k$}
\end{cases}
\\
\gamma'_{ij} =& 
\begin{cases}
-\gamma_{ij} &  \mbox{if}\quad i=k \quad \mbox{or}\quad  j=k \\
 \ \gamma_{ij} + M\, {\rm max}(0, \gamma_{ik} \gamma_{kj})\, {\rm sign}(\gamma_{kj}) & 
 \mbox{if} \quad i,j\neq k
 \end{cases}
 \\
\zeta'_i=& \begin{cases} 
-\zeta_k & \hbox{if $i=k$} \\
\zeta_i+M\, \text{max}(0, \varepsilon \gamma_{ik}) \, \zeta_k & \hbox{for  $i\ne k$},
\end{cases}
\\
N'_i=&\begin{cases}
-N_k+ M \sum_{j\neq k} N_j \, {\rm max}(0,\varepsilon \gamma_{jk}) & \hbox{if $i=k$} \\
N_i &  \hbox{if $i\neq k$}
\end{cases}
\end{split}
\ee
where $M$ is an integer defined by
\be
\label{eq:M}
M \equiv \sum_{\ell\geq 1}
 \sum_{n,s} \ell^2 \, \Omega_{n,s}(\ell \gamma_k) \ ,\qquad
\OmS(\ell\gamma_k;y;t) = \sum_{n,s } \Omega_{n,s}(\ell\gamma_k) y^n t^s\ .
\ee
These transformation laws guarantee that
\be
\gamma\equiv \sum_i N_i \gamma_i = \sum_i N_i'\gamma_i'\, .
\ee

We conjecture that the Laurent polynomials produced by the Coulomb branch formula are invariant
under the generalized mutation transformation: \footnote{The second equation in \eqref{mutinv0}  may be surprising at first, but
physically it 
reflects the fact that in the transformed quiver states with 
charge vectors $\ell\gamma_k$ are considered as anti-BPS states and are no longer counted
in the BPS index. On the other hand states with charge vector $-\ell\gamma_k$, which are
considered anti-BPS in the original quiver and not counted, are taken to be BPS in the new
quiver.}
\be
\label{mutinv0}
\QC(\gamma; \zeta; y; t) =
\begin{cases} \QC'(\gamma; \zeta'; y; t)  &\hbox{if $\gamma\not\parallel \gamma_k$}\\
\QC'(-\gamma;\zeta;y;t)  &\hbox{if $\gamma\parallel \gamma_k$}\,,
\end{cases}
\ee
under the conditions that 
\begin{itemize}
\item[i)] $\Omega_{n,s}(\ell\gamma_k)$ are positive
integers satisfying $\Omega_{n,s}(\ell\gamma_k)=\Omega_{-n,-s}(\ell\gamma_k)$ and 
vanish 
for $\ell$ large enough, so that the integer $M$ is well defined, 
 \label{condi}
\be \label{eposcon0}
\Omega_{n,s}(\ell\gamma_k)\geq 0\ \forall\  \ell>0\ ,\quad
\Omega_{n,s}(\ell\gamma_k) = 0 \, \, \text{for}
\, \, \ell > \ell_{\rm Max}\, ,
\ee
\item[ii)] 
the stability parameter $\zeta_k$ has sign $-\varepsilon$,
\be
\label{einequal}
\varepsilon\, \zeta_k<0\ ,
\ee
\item[iii)] the single-centered indices transform
as\footnote{It is easy to verify that the rational invariants $\bQC$ and $\bOmS$ satisfy
the same mutation transformation rules as $\QC$ and $\OmS$ respectively.}
\be
\label{egench00}
\OmS(\alpha;y;t) =
\begin{cases}
 \OmS'\left(\alpha+ M {\rm max}(0, \varepsilon \langle \alpha,\gamma_k\rangle)\, \gamma_k
; y;t\right) & \hbox{for $\alpha \not\parallel \gamma_k$} \\
\OmS'\left( - \alpha\right) & \hbox{for $\alpha\parallel \gamma_k$}
\end{cases}\, .
\ee
\end{itemize}
In \eqref{mutinv0}, it is understood that in computing the l.h.s. we have to express
$\gamma$ as $\sum_iN_i\gamma_i$  treating $\gamma_i$'s as the basis vectors and apply
the Coulomb branch formula \eqref{essp1}, \eqref{essp2} while in 
computing the r.h.s. we have to express
$\gamma$ as $\sum_i N'_i\gamma'_i$ treating $\gamma'_i$'s as the basis vectors and then
apply the Coulomb branch formula \eqref{essp1}, \eqref{essp2}.
Since  the left and right mutations $\mu_k^\pm$ are inverses of each other, we shall
restrict our attention to right mutations only and set  
\be  
\varepsilon=1
\ee
henceforth.

Several remarks about our generalized mutation conjecture are in order:
\begin{enumerate}
\item 
For ordinary quivers,  
$\OmS(\ell\gamma_k)=\delta_{\ell,1}$,
hence $M=1$ and 
the above relations reduce to mutations of ordinary quivers with superpotential (the
action on the superpotential can be found in \cite{zbMATH05573998}). 
\item For quivers 
obtained from cluster algebras with  skew-symmetrizable exchange matrix (i.e. a integer matrix
 $\hat \gamma_{ij}$ such that $\gamma_{ij}\equiv \hat\gamma_{ij}/d_j$ is antisymmetric for some positive integers $d_i$), the action on $\gamma_{ij}$ coincides with the mutation rule specified in 
\cite{2003math11245F,zbMATH05145743} for $M=d_k$.

\item Mutation invariance in general imposes additional restrictions
on the single-centered invariants $\OmS(\gamma;y;t)$, beyond the vanishing of  $\OmS(\gamma;y;t)$ for $\gamma\notin\Gamma^+$ with respect to the original quiver. 
Indeed, if we denote  by $\Gamma'^+$
the set of vectors $\gamma=\sum_i n_i'\gamma_i' \in\Gamma$ where all $n_i'$ are non-negative,
then the transformation rule \eqref{egench00} requires that $\OmS(\alpha;y;t)$ should
vanish if the mutated vector $\alpha'\equiv \alpha+ M {\rm max}(0, \langle \alpha,\gamma_k\rangle)\, \gamma_k$ does not lie in  $\Gamma'^+$, even if $\alpha\in\Gamma^+$  (excluding 
the case $\alpha\parallel\gamma_k$)\footnote{The same reasoning applies to the Dolbeault-Poincar\'e polynomial: $\QC(\gamma)=0$ if $\gamma\notin \Gamma_+$ or $\gamma\notin\Gamma'_+$}. 
Similarly, $\OmS'(\alpha')$ should vanish if $\alpha'\in \Gamma'^+$ but $\alpha\notin\Gamma^+$. Another consequence of the generalized mutation symmetry is that $\OmS(\gamma_j + \ell\gamma_k)$ must vanish for all $\ell\ne 0$. Indeed, for negative $\ell$, the vector $\alpha=\gamma_j + \ell\gamma_k$ fails to lie in $\Gamma^+$,
while for positive $\ell$, the mutated vector $\alpha'=\gamma_j + M {\rm max}(\gamma_{jk},0)\, \gamma_k+\ell\gamma_k=\gamma_j' - \ell\gamma'_k$ fails to lie in $\Gamma'^+$. 
If the $\Omega_s$'s fail to satisfy these constraints, they still define 
a generalized quiver but generalized mutation symmetry does not apply. 
Indeed
it is unclear a priori if there exists a set 
of single-centered invariants $\OmS(\gamma;y;t)$ which is consistent with the above  
constraints arising from  arbitrary sequences of mutations. Finding a Higgs branch-type realization of such generalized quivers invariant under mutations would allow to give an affirmative 
answer to this question. 
\item A useful way to state the property \eqref{mutinv0} 
is to construct the  generating functions
\ben
\label{defFF}
\FF(\vec N; \zeta; q; y; t)  &\equiv& \sum_{N_k} \QC\left(\sum_{i\ne k} N_i \gamma_i
+ N_k \gamma_k; \zeta; y; t\right) q^{N_k} \, , \non \\
\FF'(\vec N; \zeta'; q; y; t)  &\equiv& \sum_{N_k} \QC'\left(\sum_{i\ne k} N_i \gamma_i'
+ N_k \gamma_k'; \zeta'; y; t\right) q^{N_k} \, ,
\een
where, on the left-hand side, $\vec N$ denotes the truncated dimension vector
\be
\vec N \equiv (N_1, \cdots N_{k-2}, N_{k-1}, N_{k+1}, N_{k+2}, \cdots )\ .
\ee
Mutation invariance for all values of $N_k$ is then equivalent to the functional identity
\be
 \FF(\vec N; \zeta; q; y; t)  = 
 \begin{cases} q^{\sum_{i\ne k} M N_i 
{\rm max}(\gamma_{ik},0)} \FF'(\vec N; \zeta'; q^{-1}; y; t)  &  \text{for}\  \vec N\ne \vec 0 \\
 \FF'(\vec 0; \zeta'; q; y; t) &  \text{for}\  \vec N=\vec 0
 \end{cases} 
\ee
We conjecture that under the assumption \eqref{eposcon0}, both sides of this equation
are in fact polynomials in $q$. 

\item
While the conditions i) -- iii) are necessary  for mutation invariance 
of the  Dolbeault polynomials $\QC(\gamma;\zeta;y;t)$,
it is possible to relax condition i)  if one is interested only 
in the numerical invariants $\QC(\gamma;\zeta;y=1;t=1)$. 
In that case we conjecture that it is sufficient that the 
generating function  $\FF(\vec N;\zeta;q;1;1)$  be a polynomial in
$q$, invariant under $q\to 1/q$ (up to an overall power
$q^{\sum_{j\neq k} M {\rm max}(\gamma_{jk},0)}$). This allows some of
the $\OmS(\ell\gamma_k;1;1)$'s to be negative.
For example, for
the generalized Kronecker quiver  (example 1 in \S\ref{sgen}), 
one may take  $\OmS(\gamma_k;1;1)=-1$,
$\OmS(2\gamma_k;1;1)=1$, and $\OmS(\ell\gamma_k;1;1)=0$ for all other
$\ell$. Then the generalized mutation $\mu_2^+$ has $M=3$ and  preserves the
numerical invariants $Q(\gamma;\zeta;1;1)$. 
Example 2(g) of \S\ref{sgen} gives another example of this phenomenon for a three-node
quiver. 

\end{enumerate}

Although we do not have a general proof that the Coulomb branch formula is indeed
invariant under such generalized mutations, we shall check 
it in many examples of
ordinary and generalized quivers, with or without oriented loop. In some cases, mutation
invariance allows to determine the complete set of single-centered
indices. Another useful property of mutations is
that in special cases they can reduce the total rank of the quiver, which typically reduces
considerably the computation time of the Coulomb branch formula.

\subsection{Outline}

The rest of the paper is organised as follows. 
In \S\ref{sgenphys} we describe the physical origin 
of the generalized
mutation transformation rules, 
the transformation properties of
single-centered indices under generalized mutation 
and the choice of FI
parameters given in \eqref{einequal}. In   
\S\ref{sord} we test the ordinary
mutation symmetry of the Coulomb branch formula through several
examples. In \S\ref{sgen} we repeat this exercise for generalized mutations.

\section{Motivation for the generalized mutation conjecture} \label{sgenphys}

\begin{figure}
\centerline{\includegraphics[height=10cm]{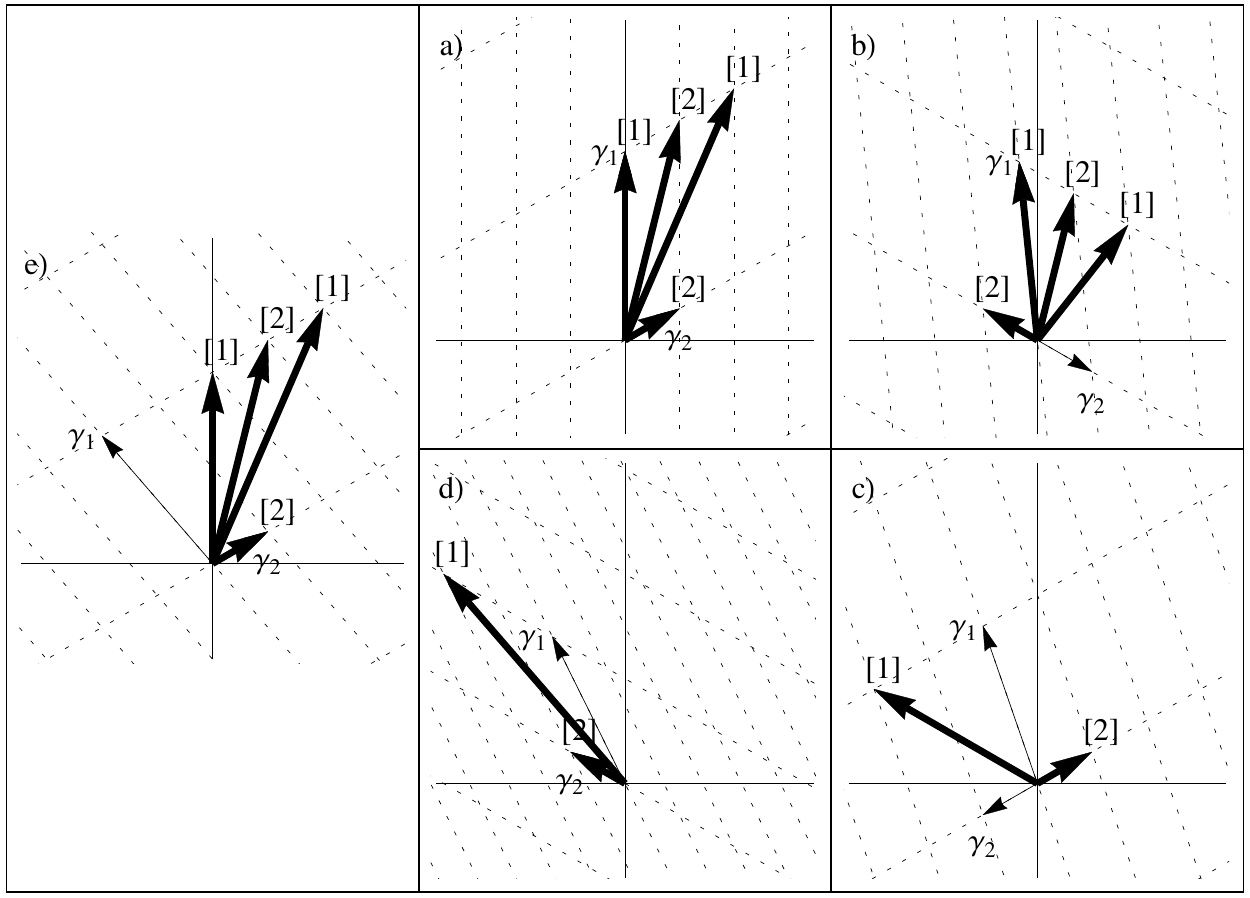}}
 \caption{Spectrum of a generalized Kronecker quiver with $\gamma_{12}=1$,
$\OmS(\gamma_1)=1, \OmS(\gamma_2)=2$ as the central charge
$Z(\gamma_2)=\rho\, e^{i\theta}$ rotates clockwise around 0, keeping
$0<\rho\ll 1$  and $Z(\gamma_1)+\tfrac{1}{\pi}(\tfrac{\pi}{6}-\theta) Z(\gamma_2)=e^{i\pi/2}$
fixed. The BPS half-space ${\rm Im}(Z)>0$ is kept fixed during the deformation. 
Occupied charges  
are depicted by an arrow in the central charge plane, decorated with
the corresponding BPS index in square bracket. 
A conjugation wall is crossed in going from a) to b) and c) to d), while walls of marginal
stability are crossed in going from b) to c) and d) to e). The spectrum in e) is identical to the
spectrum in a), up to a monodromy $\gamma_1\mapsto\gamma_1+2\gamma_2$. In more detail:
a) $0<\theta<\pi/2$: the spectrum consists of 4 occupied charges 
$(\gamma_1,\gamma_1+\gamma_2,\gamma_1+2\gamma_2,\gamma_2)$ and
BPS indices $(1,2,1,2)$, respectively.  b) $-\pi/2<\theta<0$: $\gamma_2$
is now anti-BPS. The spectrum of the mutated quiver  consists of 4
occupied charges 
 $(-\gamma_2,\gamma_1,\gamma_1+\gamma_2,\gamma_1+2\gamma_2)$
and indices $(2,1,2,1)$.  c)  $-\pi<\theta<-\pi/2$:  the phases of
the  two charges 
$(\gamma_1+2\gamma_2,-\gamma_2)$ swap and they no
longer form any BPS bound state. d) $-3\pi/2<\theta<-\pi$: $\gamma_2$
re-enters the BPS half space and the spectrum of the twice-mutated
quiver contains two occupied charges 
$(\gamma_2,\gamma_1+2\gamma_2)$ with index $(2,1)$ and no bound state. 
e) $-2\pi<\theta<-3\pi/2$, the phases of the two charges 
$(\gamma_2,\gamma_1+2\gamma_2)$ swap again and the spectrum  of the
twice-mutated quiver consists of 4 occupied charges 
$(\gamma_1+2\gamma_2,\gamma_1+3\gamma_2,\gamma_1+4\gamma_2,\gamma_2)$
with indices $(1,2,1,2)$. }\label{figB2}
\end{figure}

As mentioned in the introduction, quiver quantum mechanics describes the dynamics of open strings stretched between the various BPS constituents of a given bound state. In particular,
it depends on a choice of half-space $\cH$ in the central charge plane, such that all states whose
central charge lie in $\cH$ are deemed to be BPS, while those in the opposite half-plane are anti-BPS. As the choice of $\cH$ is varied, it may happen that one of the constituents, with charge $\gamma_k$, crosses the boundary of $\cH$ and falls on the anti-BPS side, while its CPT-conjugate with charge $-\gamma_k$ enters the BPS side.\footnote{We assume that the spectrum is such that no other BPS
state crosses the boundary of $\cH$ at the same time.} Equivalently, this may take place
for a fixed choice of $\cH$ under a variation of the asymptotic moduli (staying away from walls of marginal stability). Such a wall is sometimes known as a wall of second kind \cite{ks}, or as a
conjugation wall \cite{Andriyash:2010yf}. Such walls are encountered in particular when varying the moduli around a point where the central charge associated to one of the BPS constituents vanishes, see Figure \ref{figB2} for an example which can serve as a guidance
for the discussion below.

Clearly, as the state with charge $-\gamma_k$ enters the BPS half-space, it cannot be viewed as a bound state of the BPS constituents with charges $\gamma_i$, and must therefore be considered as elementary. Consequently the vector  $-\gamma_k$  must be taken as a new basis 
vector, and   the other basis vectors must be changed as well so that 
the charges carried by the BPS states can be expressed as positive linear combinations of the
basis vectors. Invariance under mutation is the statement that the same 
BPS states can be described either as bound states of the original BPS constituents with charge $\gamma_i$,
or of the new BPS constituents with charge $\gamma'_i$.   

For this equivalence to hold, it is not necessary  
that the indices associated with  
the constituents satisfy the constraint \eqref{OmSquivers} -- 
indeed this constraint is generically not obeyed for bound states in gauge theory 
\cite[Section 3.2]{Cordova:2013bza}  and in supergravity (such as in 
the D6-D0 system, studied in more detail in \cite[Appendix B]{Manschot:2010qz}). 
Instead, we shall allow the indices $\OmS(\gamma_j)$ of the BPS constituents to be arbitrary symmetric Laurent polynomials in two-parameter  $y$ and $t$, with support on non-negative
dimension vectors $\gamma\in\Gamma^+$. 
We refer to the polynomials 
$\QC(\gamma;\zeta;y;t)$ produced by the Coulomb branch formula \eqref{essp1} as generalized quiver invariants. We also assume that 
 $\OmS(\gamma_j+\ell \gamma_k)$ vanishes for $\ell\ge 1$, and that the integers
$\Omega_{n,s}(\ell\gamma_k)$  defined through \eqref{eq:M}
are all positive and vanish  for some large enough $\ell$.
The necessity of the first   
 condition was discussed in the last but one paragraph of \S\ref{sgenmut}, whereas the
 necessity of the 
second  condition will become clear below. 
Figure \ref{figB2} is an example of  generalized quivers, associated to a rank 2 cluster algebra 
with non-symmetrizable exchange matrix with Dynkin diagram $B_2$ (see \cite{Alexandrov:2011ac} for a similar example with Dynkin diagram $G_2$).
In the rest of this section
we shall describe the motivation behind
 the generalized mutation conjecture
\eqref{mutDSZgen0}-\eqref{egench00}
for the generalized quiver invariants.

\subsection{Semi-primitive Coulomb formula and Fermi flip} \label{secflip}

In order to motivate the action of mutations on the basis of BPS states, we shall focus on
dimension vectors $\gamma= \, \gamma_j + N\gamma_k$   
with support only on two nodes, the mutating node $k$ and any adjacent
node $j$, hence effectively dealing with a Kronecker quiver with $\gamma_{jk}$ arrows
and dimension vector $(1,N)$.

Due to our assumption that 
 $\OmS(\gamma_j+\ell \gamma_k)=0$ for non-zero $\ell$,  states 
 carrying charge $\gamma_j+N\, \gamma_k$ 
can only arise in the original quiver 
as bound states of a center of charge $\gamma_j$ with other centers carrying
charges $\ell_i\gamma_k$ with $\ell_i>0$. 
Assuming $\zeta_k<0<\zeta_j$, these states exist whenever 
$\gamma_{jk}>0$, and arise physically as halos of particles of charge $\ell\gamma_k$ orbiting around a core of charge $\gamma_j$ \cite{Denef:2007vg}. Their
indices are given by the semi-primitive Coulomb branch 
formula\cite{Denef:2007vg,Dimofte:2009bv,Manschot:2010qz}, 
\be
\label{eqor}
\begin{split}
Z=&\sum_N \QC(\gamma_j+N \gamma_k;\zeta;y;t) \, q^N \\
=& \OmS(\gamma_j;y;t) \, 
\prod_{\ell\ge 1} \prod_{J=1}^{\ell\gamma_{jk}}\prod_n \prod_s \left( 1 + q^\ell t^s y^n (-y)^{2J - \ell \gamma_{jk}
-1}\right)^{\Omega_{n,s}(\ell\gamma_k)} \, .
\end{split}
\ee
This implies that only a finite number of charge vectors $\gamma_j+N \gamma_k$ have non-zero index, namely those with $0\leq N\leq M\, \gamma_{jk}$ where
\be
\label{defM}
M \equiv \sum_{\ell= 1}^{\ell_{\rm Max}} \sum_{n,s} \ell^2 \, \Omega_{n,s}(\ell \gamma_k) \, .
\ee
Physically $\QC(\gamma_j+N \gamma_k;\zeta;y;t)$ can be interpreted as the number of states
corresponding to the excitations of the fermionic oscillators of 
charges $\ell_i\gamma_k$   
in \eqref{eqor} acting on the fermionic vacuum with charge $\gamma_j$. As pointed out in
\cite{Andriyash:2010yf}, the same multiplet of states can be obtained from the filled Fermi sea
of charge $\gamma'_j=\gamma_j+M \gamma_{jk} \gamma_k$ by acting with fermionic oscillators of  charges $\ell_i\gamma'_k=-\ell_i\gamma_k$, 
provided they carry the same  indices 
 \be \label{esemireln}
 \Omega'_{n,s}(\ell \gamma'_k) = \Omega_{n,s}(\ell \gamma_k)\ ,\qquad
 \OmS'(\gamma_j';y;t) =\OmS(\gamma_j;y;t)  \ .
 \ee
 The particles of charge $\ell\gamma'_k$ and $\gamma'_j$ and the corresponding
 indices can be associated to the nodes of a new
 (generalized) quiver. 
 In this alternative description,  
 the bound states with charge $\gamma_j+N \gamma_k=\gamma_j'+(M\gamma_{jk}-N) \gamma_k'$
are described in terms of a halo of particles of charges $\ell_i\gamma'_k$ orbiting around a 
core of charge $\gamma'_j$. To see the equivalence of the two descriptions, 
one can start from the halo partition function 
 \ben
  \label{eqor1}
Z'&\equiv& \sum_N \QC'(\gamma_j'+(M  \gamma_{jk} -N) \gamma_k';\zeta';y;t) \, q^N \non\\
&=& q^{M\gamma_{jk} } \sum_{N'} \QC'(\gamma_j'+N' \gamma_k';\zeta';y;t) \, q^{-N'} \non\\
&=& q^{M\gamma_{jk} } \, \OmS'(\gamma_j';y;t) \, 
\prod_{\ell\ge 1} \prod_{J=1}^{\ell\gamma_{jk}}\prod_n \prod_s \left( 1 + q^{-\ell} t^s y^n (-y)^{2J - \ell \gamma_{jk}
-1}\right)^{\Omega_{n,s}'(\ell\gamma_k')}\, ,
\een
 where we have used the fact that 
$\gamma'_{jk} = -\gamma_{jk}<0$ and $\zeta'_k>0$. Taking out the
factor of $ q^{-\ell} t^s y^n (-y)^{2J - \ell \gamma_{jk}
-1}$ from each term inside the product in \eqref{eqor1}, using \eqref{esemireln}
and making a change of variable $J\to \ell\gamma_{jk}-J+1$,
this can be rewritten as
\ben
Z'&=& q^{M\gamma_{jk} - \gamma_{jk} \sum_{\ell,n,s} \ell^2 \Omega_{n,s}(\ell\gamma_k)} \, 
t^{\gamma_{jk} \sum_{\ell,n,s} \ell\, s\,  \Omega_{n,s}(\ell\gamma_k)}
y^{\gamma_{jk} \sum_{\ell,n,s} \ell\, n\,  \Omega_{n,s}(\ell\gamma_k)}\non \\
&&
\OmS(\gamma_j;y;t) \, 
\prod_{\ell\ge 1} \prod_{J=1}^{\ell\gamma_{jk}}\prod_n \prod_s \left( 1 + q^{\ell} t^{-s} y^{-n} 
(-y)^{2J - \ell \gamma_{jk}
-1}\right)^{\Omega_{n,s}(\ell\gamma_k)}\, .
\een
The exponent of $q$ in the first factor on the right hand side vanishes due to \eqref{defM},
while the exponents of $t$ and $y$ in the second and third factors vanish due to the
Hodge duality symmetry $\Omega_{n,s}(\ell\gamma_k) = \Omega_{-n,-s}(\ell\gamma_k)$.
The same symmetry allows us to replace the $t^{-s} y^{-n}$ term inside the product by
$t^{s} y^{n}$. Thus we arrive at
\be
Z'=
\OmS(\gamma_j;y;t) \, 
\prod_{\ell\ge 1} \prod_{J=1}^{\ell\gamma_{jk}}\prod_n \prod_s \left( 1 + q^{\ell} t^{s} y^{n} 
(-y)^{2J - \ell \gamma_{jk}
-1}\right)^{\Omega_{n,s}(\ell\gamma_k)}\, ,
\ee
reproducing \eqref{eqor} whenever $\gamma_{jk}>0$. 
If  instead $\gamma_{jk}<0$ (keeping $\zeta_k<0<\zeta_j$) then  the first quiver does not 
carry any bound state of the center carrying charge $\gamma_j$ with centers carrying
charges $\ell_i\gamma_k$ with $\ell_i>0$. Thus $\QC(\gamma_j+N\gamma_k)$ vanishes
for $N>0$. The mutated quiver describing 
 centers of charges $\gamma_j'=\gamma_j$
and $\ell_i\gamma_k'=-\ell_i\gamma_k$, with indices $\OmS(\gamma_j;y;t)$ and
$\OmS(\ell_i\gamma_k;y;t)$ respectively, has
 $\gamma'_{jk}>0$, $\zeta_j'<0<\zeta'_k$,
and therefore also no bound states  of charge $\gamma_j'+N\gamma_k'$ for $N>0$.
The partition functions $Z=Z'=\OmS(\gamma_j;y;t)$ are therefore again the same on both sides. 

This shows that, under the assumptions $\zeta_k<0< \zeta_j$  and \eqref{eposcon0}, the 
semi-primitive Coulomb branch formula is invariant under the transformation
\ben \label{esemigen}
&& \gamma_k'= -\gamma_k\ ,\quad
\gamma_j'=\gamma_j+M\, \text{max}(0, \gamma_{jk})\, \gamma_k \quad \hbox{for $j\ne k$}, 
\nonumber \\
&& \OmS(\gamma_j) = \OmS'(\gamma_j'), \quad
\OmS'(\ell\gamma'_k; y; t) = \OmS(\ell \gamma_k; y; t) \, \quad \forall \ell
\, .
\een
This is  a special case of the generalized mutation rules  \eqref{mutDSZgen0}-\eqref{egench00},
providing the initial motivation for the conjectured invariance under the
generalized mutation transformation. 
In the next subsections, we comment on
aspects of the generalized mutation rules which are not obvious consequences of 
the semi-primitive case.

\subsection{Transformation rule of single-centered indices}

Let us now comment on the transformation rule \eqref{egench00}  of
$\OmS(\alpha)$. The first equation for $\alpha=\gamma_j$ as well as the second equation
follow from the analysis of the Kronecker quiver given
above,\footnote{While this paper was in preparation, this observation
  was also made in Ref. \cite{Cordova:2013bza}.} but we shall now justify why this is needed for
general $\alpha$.
Consider two generalized
quivers which are identical in all respects except that for some specific charge vector
$\alpha$, the first quiver has $\OmS(\alpha)=0$ while the second quiver has some 
non-zero $\OmS(\alpha;y;t)$. Let us denote by $Q(\gamma)$ and $\hat Q(\gamma)$ the
Coulomb branch formul\ae\ for these two quivers. Now consider the difference
$\hat Q(\alpha+\ell\gamma_k) - Q(\alpha+\ell\gamma_k)$ for some positive integer $\ell$.
This difference must come from a bound state configuration of a center of charge
$\alpha$ with a set
of centers carrying charges parallel to $\gamma_k$. The index associated with this
configuration is encoded in the partition function $Z$ given in \eqref{eqor} with 
$\gamma_j$ replaced by $\alpha$. Now consider the mutated version of both quivers with
respect to the $k$-th node. The difference 
$\hat Q'(\alpha+\ell\gamma_k) - Q'(\alpha+\ell\gamma_k)$ must agree with
$\hat Q(\alpha+\ell\gamma_k) - Q(\alpha+\ell\gamma_k)$. Our previous analysis showing the
equality of $Z$ and $Z'$ guarantees
that this is achieved if we assume that  the mutated quivers are identical
except for one change:
$\OmS'\left(\alpha+ M {\rm max}(0, \langle \alpha,\gamma_k\rangle)\, \gamma_k
; y;t\right)$ is zero in the first mutated quiver but is
equal to $\OmS(\alpha;y;t)$ for the second mutated quiver. The extra states in the second 
quiver then appear from the bound state of a center carrying charge 
$\alpha+ M\, \gamma_k\, {\rm max}(0, \langle \alpha,\gamma_k\rangle)$ 
and other states with charges
proportional to $-\gamma_k$.
This in turn justifies the transformation law of $\OmS$  given in the  
first equation of \eqref{egench00}. 

This transformation law is also consistent with the requirement that a monodromy, 
exemplified in Figure \ref{figB2},
leaves invariant the physical properties of the BPS spectrum. 
Since the monodromy transformation is induced by successive
application of two mutations, one with a node carrying charge proportional to $\gamma_k$ and
then with a node carrying charges proportional to $-\gamma_k$, the 
transformation law \eqref{egench00} under a mutation implies that under a monodromy
we have $\tilde \Omega_{\rm
  S}(\alpha+M\langle\alpha,\gamma_k\rangle \gamma_k)=\OmS(\alpha)$, where we denoted
 by $\tilde \Omega_{\rm
  S}$ the single centered indices after the monodromy transformation. On the other hand 
a monodromy maps a BPS bound state with constituent
charges $\alpha$ to one with charges $\tilde
\alpha=\alpha+M\langle\alpha,\gamma_k\rangle\, \gamma_k$,  while other physical
quantities as the central charges and symplectic inner products remain
invariant. Moreover, the physical equivalence of the bound states
before and after the monodromy requires that the single centered
indices transform as $\tilde \Omega_{\rm
  S}(\tilde\alpha)=\OmS(\alpha)$.
This agrees with the monodromy transformation law of $\OmS$
obtained by application of two successive mutations.

\subsection{Dependence on the choice of FI parameters} \label{sfichoice}

Note that while \eqref{einequal} fixes the sign of $\zeta_k$, it leaves
unfixed the signs and the magnitudes of the other $\zeta_i$'s as long as they
satisfy $\sum_i N_i\zeta_i=0$. Since for different choices of the FI parameters
we have different $\QC$ and $\QC'$, \eqref{mutinv0} apparently gives different
consistency relations for different choices of FI parameters. We shall now outline
a proof that once the mutation invariance has been tested for one choice
of FI parameters, its validity for other choices of FI parameters subject to the
restriction \eqref{einequal} is automatic. 
We shall carry out the proof in steps.

First consider a vector $\gamma\in\Gamma^+\backslash \Gamma'^+$ (i.e. such that $\gamma=\sum_i n_i \gamma_i=\sum_i n_i'\gamma_i'$ with
non-negative $n_i$'s, but  with some negative $n_k'$). 
 In this case $Q'(\gamma)$ (and the rational invariant 
$\bar Q'(\gamma)$) vanishes in all chambers and hence $Q(\gamma)$ and
$\bar Q(\gamma)$  must
also vanish in all chambers. 
We shall now prove that it is enough to check that
$\bar Q(\gamma)$ vanishes in any one chamber, by induction
on the rank $r=\sum n_i$.\footnote{Note that  the
rank depends on whether we are using the original or the mutated quiver. 
Here rank will refer to the rank in the original quiver.}
Suppose that we have
verified the vanishing of $\bar Q(\gamma)$ for all $\gamma\in \Gamma^+\backslash \Gamma'^+$ with rank $\le r_0$
for some integer $r_0$.
Now consider a $\gamma\in 
\Gamma^+\backslash \Gamma'^+$ with rank $r=r_0+1$, and suppose
that $\bar Q(\gamma)$ vanishes in some chamber $c_+$.  If we now go across a wall of
$c_+$ then the jump in $\bar Q(\gamma)$ across the wall will be given by the sum of products
of $\bar Q(\alpha_i)$ for appropriate charge vectors $\alpha_i$ satisfying 
$\sum_i\alpha_i=\gamma$. Now in the original quiver each of the $\alpha_i$'s have
rank less than 
$r_0$. Furthermore at least one of the $\alpha_i$'s  must be in
$\gamma\in \Gamma^+\backslash \Gamma'^+$; to see this note that when we express $\gamma=\sum_i \alpha_i$ 
in the $\gamma'_i$ basis
the coefficient of $\gamma'_k$ is negative, and hence at least one of the $\alpha_i$'s
expressed in the $\gamma'_i$ basis has negative coefficient of $\gamma'_k$. Thus the
corresponding $\bar Q(\alpha_i)$ vanishes by assumption, causing the net jump in 
$\bar Q(\gamma)$ to vanish. Thus the vanishing of $\bar Q(\gamma)$ in one chamber
implies its vanishing in all chambers.
Similarly, if $\gamma\in\Gamma'^+\backslash \Gamma^+$, the same argument shows that 
the vanishing of $Q'(\gamma)$ in one chamber is sufficient to
ensure the vanishing in all chambers.

Now suppose that we have already established the vanishing of $Q(\gamma)$ for 
$\gamma\in \Gamma^+\backslash \Gamma'^+$ and of $Q'(\gamma)$ for $\gamma\in \Gamma'^+\backslash \Gamma^+$ in all the chambers
subject to the restriction
\eqref{einequal}.
We now consider a general charge vector $\gamma$. 
Our goal will be to show that to test the equivalence of $Q(\gamma)$ and
$Q'(\gamma)$, it is enough to verify this in one chamber for each $\gamma$.
We shall carry out this proof by induction.
Let us suppose that 
we have
established the equality of $Q(\gamma)$ and $Q'(\gamma)$ for all $\gamma$ 
(except for $\gamma\parallel \gamma_k$)
of
rank $\le r_0$ in the $\gamma_i$ basis 
in all chambers subject to the restriction
\eqref{einequal}.
We shall then prove that for a 
charge vector $\gamma$ of rank $r_0+1$, the  equality of
$Q(\gamma)$ and $Q'(\gamma)$ in any one chamber $c_+$ implies their 
equality in all chambers.
For this consider a wall of marginal stability that forms a boundary of $c_+$.
Then as we approach this wall we can find a pair of primitive charge vectors
$\alpha_1$ and $\alpha_2$ such that $\gamma=M_1\alpha_1+M_2\alpha_2$ for
positive integer $M_1$ and $M_2$ and furthermore the FI parameters associated
with the vectors $\alpha_1$ and $\alpha_2$ 
change sign across the wall. 
Using the wall-crossing formula, the jump in $Q(\gamma)$ across the wall
can be expressed as a sum of products 
of $\bar Q(m\alpha_1+n\alpha_2)$ for integer $m,n$ in appropriate chambers relevant for
those quivers. 
Similarly the jump 
in $Q'(\gamma)$ can be expressed as a sum of products of $Q'(m\alpha_1+n\alpha_2)$
for positive integer $m,n$ in the same chambers using the same wall-crossing formula. 
Now since $m\alpha_1 + n\alpha_2$, being a constituent of the charge vector $\gamma$,
must have rank $<r_0$ 
in the original quiver, the equality of $Q(m\alpha_1+n\alpha_2)$
and $Q'(m\alpha_1+n\alpha_2)$ in any chamber holds by assumption. This shows that the net jumps
in $Q(\gamma)$ and $Q'(\gamma)$ across the wall agree and hence $Q(\gamma)=
Q'(\gamma)$ on the other side of the wall. 

There are two possible caveats in this argument. First we have to assume that 
none of the constituents
carrying charge $m\alpha_1+n\alpha_2$
has charge proportional to $\gamma_k$ since the equality of $Q(\gamma)$ and
$Q'(\gamma)$ does not hold for these charge vectors. This is guaranteed as long as we
do not cross the $\zeta_k=0$ wall, \i.e.\ as long as we obey the constraint \eqref{einequal}.
Second, we have implicitly assumed that for every possible set
of constituents\footnote{Here by constituent we do not mean only single-centered  
constituents but also bound systems whose single-centered
constituents remain at finite separation
as we approach the wall. The index carried by such a constituent of charge $\alpha$ is
given by $Q(\alpha)$ in appropriate chamber.}
in the first quiver there is a corresponding set of constituents in the 
second quiver carrying the same index 
and vice versa. This is not true in general since there may be constituents in the first
quiver whose image in the second quiver may contain one of more $\alpha_i$'s with negative
coefficient of $\gamma_k'$ and hence is not a part of the second quiver. 
These are the $\alpha_i$'s  belonging to $\Gamma^+\backslash \Gamma'^+$.
The reverse is also possible. However since we have assumed
that the vanishing of  $Q(\alpha_i)=0$ for all $\alpha_i\in \Gamma^+\backslash \Gamma'^+$ and 
the vanishing of  $Q'(\alpha_i)=0$ for all $\alpha_i\in \Gamma'^+\backslash \Gamma^+$
has already been established, 
these possible non-matching contributions vanish identically
and we get the equality of $Q(\gamma)$ and $Q'(\gamma)$ in all chambers.
This establishes that, for any $\gamma\in\Gamma$,  
the equality of $Q(\gamma)$ and $Q'(\gamma)$ in all chambers
follows from the equality in any given chamber.

We end by giving a physical motivation for 
the restriction on the FI parameters
given in \eqref{einequal}. As explained earlier, in 
$\NN=2$ supersymmetric theories where quiver invariants capture the index
of BPS states,  the mutation $\mu^+_k$  takes place on walls 
where the central charge $Z(\gamma_k)$ leaves the half-plane distinguishing 
BPS states from anti-BPS states, while $Z(-\gamma_k)$ enters the same half-plane.
This clearly
requires that in the complex plane the ray of $Z(\gamma_k)$ lies to the extreme left of the
ray of any other $Z(\gamma)$ inside the BPS half-plane. Now the FI
parameter associated with $\gamma_k$ for a particular quiver of total charge $\gamma$ is
given by 
\be 
\zeta_k = \text{Im} (Z(\gamma_k) / Z(\gamma)) \, .
\ee
The condition on $Z(\gamma_k)$ mentioned above requires that $\zeta_k$ is negative.
However it does not specify its magnitude, nor the 
magnitude or signs of the $\zeta_i$'s carried the other 
constituents, as those depend both on the phases of $Z(\gamma_i)$ and on their 
magnitudes. Thus we see from this physical consideration that if mutation is to be a symmetry,
it must hold under the condition \eqref{einequal} with no further constraint on the 
other $\zeta_i$'s.

\section{Examples of ordinary quiver mutations} \label{sord}

In this section we shall test mutation invariance of the Coulomb branch formula
for ordinary quivers. 
For this we take $\OmS(\gamma)$ to satisfy \eqref{OmSquivers}
and use the transformation law \eqref{egench00} of  $\OmS(\gamma)$ under mutation.
We also 
use mutation invariance to 
compute single-centered indices for various quivers where a 
direct analysis of the Higgs branch is forbidding. Since ordinary mutation is known
to be a symmetry of the quiver Poincar\'e polynomial, the analysis of this section can be
interpreted as a test of the Coulomb branch formula \eqref{essp1}, \eqref{essp2} and the
transformation rule \eqref{egench00} for single-centered indices.

\bigskip

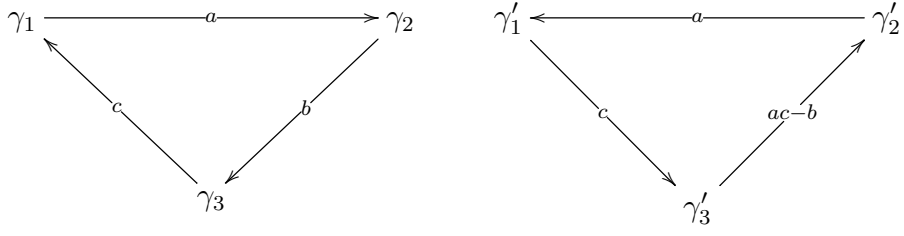
\begin{figure}
\begin{center}
$$
\xymatrix{
\gamma_1 \ar[rrrr]|{a}  & & & &  \gamma_2 \ar[lldd]|{b} \\  &&&& \\
& & \gamma_3 \ar[lluu]|{c} & & } \qquad
\xymatrix{
\gamma_1' \ar[rrdd]|{c}  & & & &  \gamma_2' \ar[llll]|{a} \\  &&&& \\
& & \gamma_3' \ar[rruu]|{ac-b} & & }
$$
\end{center}
\caption{The original quiver (left) and the mutated quiver (right) of examples 1 and 2.
\label{f1}.}
\end{figure}

\noindent {\bf Example 1}: Consider a 3-node quiver with charge vectors $\gamma_1$,
$\gamma_2$ and $\gamma_3$ associated with the nodes satisfying
\be
\gamma_{12}=a, \quad \gamma_{23} = b, \quad \gamma_{31} = c, \qquad
\zeta_1< 0, \quad \zeta_2, \zeta_3>0, \quad a,b,c>0\, .
\ee
Then mutation with respect to the node 1 generates a new quiver with basis vectors
\be \label{e3.8}
\gamma_1'=-\gamma_1, \quad \gamma_2'=\gamma_2, \quad \gamma_3'
= \gamma_3 + c\, \gamma_1,
 \ee
DSZ matrix
\be
\gamma'_{12}=-a, \quad \gamma'_{23}=b - ac, 
\quad \gamma'_{31} = -c,
\ee
FI parameters
\be
\zeta_1'= - \zeta_1, \quad \zeta_2' = \zeta_2, \quad \zeta_3'=\zeta_3 + c\, \zeta_1\, ,
\ee
and dimension vector
\be
\gamma = N_1' \gamma_1' + N_2' \gamma_2' + N_3' \gamma_3', \qquad
N_1' = c N_3  - N_1, \quad N_2' = N_2, \quad N_3' = N_3\, .
\ee
The original and mutated quiver are depicted  in Fig.\ref{f1}. 

Mutation invariance \eqref{mutinv0} requires 
\be \label{eqcmut}
Q(N_1, N_2, N_3) =Q'(N_1', N_2', N_3')\ ,
\ee
where the l.h.s. is the shorthand notation for 
$\QC\left(\sum_{i=1}^3 N_i \gamma_i;\zeta; y;t\right) $ 
while the r.h.s. is the shorthand notation for 
$\QC'\left(\sum_{i=1}^3 N_i' \gamma_i';\zeta'; y;t\right) $,
computed with $\gamma_i'$ as the basis vectors and hence
$\gamma'_{ij}$ as the DSZ products.
We shall also use 
$\OmS(N_1, N_2, N_3)$ to denote $\OmS(\sum_{i=1}^3 N_i \gamma_i)$
and $\OmS'(N_1', N_2', N_3')$ to denote $\OmS'(\sum_{i=1}^3 N_i' \gamma_i';t)$.
Eq.\eqref{egench00} then gives
\be \label{eomtrs}
\OmS(N_1, N_2, N_3) = \OmS'(cN_3 - N_1 - \text{max} (0, N_3 c - N_2 a), N_2, N_3)
= \OmS'(\text{min}(N_3 c, N_2 a)-N_1, N_2, N_3)\, .
\ee

Let us choose
\be 
a=3, \quad b=4, \quad c=5, \quad \zeta_1 = -5.71, \quad \zeta_2=2.56 \, N_1/N_2+.01/N_2, \quad
\zeta_3=3.15 \, N_1/N_3 -.01/N_3\, .
\ee
Then we get
\ben
&& \gamma'_{12}=-3, \quad \gamma'_{23}=-11, \quad \gamma'_{31}=-5, \nonumber \\
&& \zeta_1'=5.71, \quad \zeta_2'=2.56\, N_1/N_2+.01/N_2, \quad \zeta_3'=3.15\, N_1/N_3 - 28.55
-.01/N_3\, .
\een
Some of the relations following from \eqref{eomtrs} are
\be \label{eomsome}
\OmS(N,1,1)=\OmS'(3-N,1,1), \quad \Rightarrow \quad
\OmS(N,1,1)=0 = \OmS'(N,1,1)  \quad \text{for} \quad N\ge 3\, .
\ee

We shall now check the invariance of the Coulomb branch formula under mutation.
Eq.\eqref{eqcmut} gives
\be \label{emut3node}
Q(N,1,1) = Q'(5-N,1,1) \quad \text{for} \quad 0\le N\le 5, \quad Q(N,1,1)=0 = Q'(N,1,1)\quad
\text{for} \quad N\ge 6
\, .
\ee
Now explicit evaluation gives
\ben
Q(1,1,1) &=& 1/y^4 + 2/y^2 + 3+ 2 y^2 + y^4 
+ \OmS(1,1,1), \nonumber \\
 Q'(4,1,1) &=& 
 1/y^4 + 2/y^2 + 3+ 2 y^2 + y^4 + 
\OmS'(2,1,1) 
 - (y + y^{-1}) 
\OmS'(3,1,1) 
+ \OmS'(4,1,1) \, .\nonumber \\ 
\een
Using \eqref{eomsome} we see that $Q(1,1,1)$ and $Q'(4,1,1)$ agree.
Next we compute
\ben 
Q(2,1,1) &=&
-(y^{-3} + 2 y^{-1} + 2 y + y^3) - (y^{-1} + y) \OmS(1,1,1) 
 +
   \OmS(2,1,1)\nonumber \\
Q'(3,1,1) &=& \OmS'(1,1,1) -(y^{-3} + 2 y^{-1} + 2 y + y^3)  - (y^{-1} + y)
\OmS'(2,1,1) +  \OmS'(3,1,1) \, . \nonumber \\
\een
Again using \eqref{eomsome} we see that $Q(2,1,1)$ and $Q'(3,1,1)$ agree. 
Similarly we have
\ben
Q(3,1,1) &=&
1 +  \OmS(1,1,1) +  \OmS(3,1,1)  - (y + y^{-1})  \OmS(2,1,1)\, , \nonumber \\
Q'(2,1,1) &=& 1 + \OmS'(2,1,1) - (y + y^{-1}) \OmS'(1,1,1)\, .
\een
These two agree as a consequence of \eqref{eomsome}.
We also have
\ben
Q(4,1,1) &=& \OmS(2,1,1) +  \OmS(4,1,1)
 - (y + y^{-1})  \OmS(3,1,1)
\nonumber \\
Q'(1,1,1) &=&\OmS'(1,1,1)\, ,
\een
which are in agreement as a consequence of \eqref{eomsome}.
Finally we have
\ben
Q(5,1,1) &=& \OmS(3,1,1) +\OmS(5,1,1) - (y + y^{-1}) \OmS(4,1,1)\, , \nonumber \\
Q'(0,1,1) &=&  0\, ,
\een
\ben 
Q(0,1,1) &=&  -(y^{-3} + y^{-1} + y + y^3) \, ,  \\
Q'(5,1,1) &=&   -(y^{-3} + y^{-1} + y + y^3) +\OmS'(3,1,1) - (y+y^{-1}) \OmS'(4,1,1)
+ \OmS'(5,1,1) \, . \nonumber
\een
Again these equations are in agreement due to \eqref{eomsome}.
We have not tested the vanishing of $Q(N,1,1)$ and $Q'(N,1,1)$ for $N\ge 6$
due to the increase in the computational time, but we shall test similar relations
involving other quivers later.

So far we have not used any explicit results for $\OmS$ or $\OmS'$. We
now note that $\OmS'(1,1,1)$ vanishes since the corresponding $\gamma'_{ij}$'s fail to
satisfy the triangle inequality. The single-centered index  
$\OmS(1,1,1;t)=9$ is easily
computed from the results in \cite{Bena:2012hf,Manschot:2012rx,Lee:2012naa}.
Thus we have
\be \label{eresom1}
\OmS'(1,1,1;t)=\OmS(2,1,1;t)=0,
\quad 
\OmS'(2,1,1;t)=\OmS(1,1,1;t)=9\, .  
\ee
It will be interesting to check the prediction for $\OmS'(2,1,1)$ by direct computation.
Note that in general $\OmS(\gamma)\neq \OmS'(\gamma)$.
For example $4\gamma_1'+\gamma_2'+\gamma_3'=
\gamma_1+\gamma_2+\gamma_3$
and $\OmS'(4,1,1) \neq 
\OmS(1,1,1)$.

\bigskip

\noindent{\bf Example 2:} We again consider a 3-node quiver with
\be 
a=2, \quad b=2, \quad c=2, \quad \zeta_1 = -3.1 , \quad \zeta_2= N_1/N_2+.2/N_2, \quad
\zeta_3=2.1\, N_1/N_3 -.2/N_3\, ,
\ee
and mutate with respect to the node 1.
Then we get
\ben
&& \gamma'_{12}=-2, \quad \gamma'_{23}=-2, \quad \gamma'_{31}=-2, \nonumber \\ &&
\zeta_1'=3.1, \quad \zeta_2'=N_1/N_2+.2/N_2, \quad \zeta_3'=2.1\, N_1/N_3 - 6.2 - .2/N_3\, ,
\een
\be \label{entrs}
N_1'=2 N_3-N_1, \quad N_2'=N_2, \quad N_3'=N_3\, .
\ee
Eqs.\eqref{eomtrs} give
\be \label{econs1}
\OmS(N_1, N_2, N_3) = \OmS'( \text{min}(2N_3, 2 N_2)- N_1, N_2, N_3)\, .
\ee
On the other hand since 
the new quiver is the same as the old one with the arrows reversed and
different FI parameters, and since $\OmS$ is independent of the FI parameters
we have 
\be \label{ecycle}
\OmS'(N_1 ,N_2,N_3)
= \OmS(N_3,N_2,N_1)\, .
\ee
Furthermore cyclic invariance of the quiver
implies that $\OmS(N_1,N_2,N_3)$ 
is invariant under cyclic permutations of $(N_1,N_2, N_3)$. Using these relations we
can severely constrain the values of $\OmS$.
For example we have\footnote{The fact that $\OmS(N,1,1)$ vanishes is 
consistent with the fact that in the chamber 
$\zeta_2>0,\zeta_1\to 0^-$ the moduli space is a codimension $Na$ surface in 
$\bP^{b-1}\times G(N,c)$, with dimension $1-N^2$.}
\be \label{eomform}
\OmS(N,1,1) = \OmS'(2-N,1,1)=\OmS(1,1,2-N)=\OmS(2-N,1,1) \, ,
\ee
and as a consequence
\be \label{eomform2}
\OmS(N,1,1)=0 \quad
\text{for} \quad N\ge 2\, .
\ee
More generally we get
\be \label{econs2}
\OmS(N_1, N_2,N_3) = 0 \quad \text{for} \quad N_1\ge \text{min}(2N_2, 2 N_3)\, .
\ee
Together with cyclic symmetry this implies that a necessary condition 
for getting non-vanishing $\OmS(N_1,N_2, N_3)$ is that
each $N_i$ should be strictly less than the double of each of the other two $N_i$'s.
Using cyclic symmetry we can take $N_1$ to be the largest of $(N_1, N_2, N_3)$.
The mutation rule \eqref{econs1} then equates $\OmS(N_1,N_2, N_3)$ 
to $\OmS'(N_1', N_2, N_3)=\OmS(N_3,N_2, N_1')$
with $N_1'\le N_1$. The equality sign holds only if $N_1=N_2=N_3$. Thus unless
$N_1=N_2=N_3$ we can repeatedly use mutation and cyclic symmetry to reduce the
rank of the quiver until the maximum $N_i$ becomes greater than or equal to twice
the minimum $N_i$, and then $\OmS$ vanishes by \eqref{econs2}. Thus the only non-vanishing
$\OmS$ in this case are $\OmS(N,N,N)$. 
We know from \cite{Bena:2012hf} that in the Abelian case, $\OmS(1,1,1;t)=1$. 

We now proceed to test the invariance of the Coulomb branch formula under mutation. 
From the general equation $Q(N_1,N_2,N_3)=Q(2N_3-N_1,N_2,N_3)$ that follows
from \eqref{entrs}, we get in particular
\be \label{eqform}
Q(N,1,1) = Q(2-N, 1,1) \quad \Rightarrow \quad Q(N,1,1)=0 \quad \text{for} \quad N\ge 3\, .
\ee
Explicit calculation gives
\ben
&& Q(1,1,1) = 1 + \OmS(1,1,1), \quad Q'(1,1,1)=1+\OmS'(1,1,1)\, , \nonumber \\
&& Q(2,1,1) = \OmS(2,1,1), \quad Q'(0,1,1)=0 \, , \nonumber \\
&& Q(0,1,1) = -(y+y^{-1}), \quad Q'(2,1,1) = -(y+y^{-1}) + \OmS'(2,1,1)\, , \nonumber \\
&& Q(3,1,1) = \OmS(3,1,1), \quad Q(4,1,1) = \OmS(4,1,1), \nonumber \\
&&  Q'(3,1,1) = \OmS'(3,1,1), \quad
Q'(4,1,1) = \OmS'(4,1,1) \, .
\een
These results are all consistent with \eqref{eqform} after we use eqs.\eqref{eomform}, 
\eqref{eomform2}.

More generally, for any 3-node quiver with $a,b>0$ and
$c=2$, the Abelian representation $(1,1,1)$
is mapped by a mutation on node 1 to an Abelian representation. 
We know from the analysis of $\OmS$ for $\vec N=(1,1,1)$ given 
in \cite{Bena:2012hf,Lee:2012sc,Lee:2012naa,Manschot:2012rx}, that the only
non-vanishing $\OmS$ arise for $a=b\geq 2$. 
In this case \eqref{eomtrs} gives
\be
\OmS(N,1,1) = \OmS'(2-N,1,1)\, .
\ee
In particular $\OmS(1,1,1)=\OmS'(1,1,1)$. 
On the other hand since in each of these cases
the arrow
multiplicities computed using \refb{e3.8}
are just reversed under the mutation, 
the equality of $\OmS(1,1,1)$ and $\OmS'(1,1,1)$ follows
automatically, confirming the transformation laws of $\OmS$ under
mutation.
Using this we can verify the equality 
of $Q(1,1,1)$ and $Q'(1,1,1)$.

\bigskip

\noindent{\bf Example 3:} 
Next we consider the 4-node quiver
\be\label{eq:quivrepK}
\xymatrix{
\gamma_1 \ar[rrr]|{a} & & & \gamma_2 \ar[dd]|{b} \\ 
& & & \\ \gamma_4 \ar[uu]|{d} & & &
\gamma_3 \ar[lll]|{c}}
\ee
with multiplicities of the arrows $a=5$, $b=5$, $c=2$ and $d=1$. We
choose for the FI parameters 
\be\label{efichoice}
\vec \zeta=\left(\frac{25\, N_4 + .1}{N_1}, \frac{17\, N_4+.2}{N_2},\frac{3\, N_4-.3}{N_3},-45\right)\, .
\ee 
We now perform a mutation at node 4. The mutated
quiver is:
\be\label{eq:quivrepK2}
\xymatrix{
\gamma_1' \ar[rrr]|{a} \ar[dd]|{d} & & & \gamma_2' \ar[dd]|{b} \\ 
& & & \\ \gamma'_4 \ar[rrr]|{c} & & &
\gamma'_3 \ar[llluu]|{cd}}
\ee
with 
\be
\gamma_1'=\gamma_1, \quad \gamma_2'=\gamma_2, \quad
\gamma'_3=\gamma_3+2\gamma_4, \quad \gamma_4'=-\gamma_4\, ,
\ee 
\be \label{ezetamut}
\vec\zeta' = \left(\frac{25\, N_4+.1}{N_1}, \frac{17\, N_4+.2}{N_2},\frac{3\, N_4-.3}{N_3}-90,45\right)\, ,
\ee
\be
N_1'=N_1, \quad N_2'=N_2, \quad N_3'=N_3, \quad N_4'=c\, N_3 - N_4 = 2\, N_3-N_4\, .
\ee
Note that the multiplicity $c$ is chosen such that the Abelian representation
$\vec N=(1,1,1,1)$ is mapped to the Abelian representation $\vec N'=(1,1,1,1)$. 
More generally Eq. \eqref{mutinv0} implies 
\be
Q(N_1,N_2, N_3, N_4) = Q'(N_1, N_2, N_3, c N_3 - N_4) = Q'(N_1, N_2, N_3, 2 N_3 - N_4)\, .
\ee
Thus we should have
\ben \label{ecomp}
&&  Q(1,1,1,0) = Q'(1,1,1,2), \quad 
Q(1,1,1,1) = Q'(1,1,1,1), \quad Q(1,1,1,2) = Q'(1,1,1,0), \nonumber \\
&&
Q(1,1,1,N) = 0=Q'(1,1,1,N) \quad \hbox{for \quad $N\ge 3$}  \, .
\een
In order to test this we need to first study the transformation law of $\OmS$. Eq.\eqref{egench00}
gives
\be
\begin{split}
\OmS(N_1,N_2,&N_3,N_4) = \OmS'(N_1,N_2,N_3,c N_3 - N_4 - \text{max}(c N_3 - d N_1, 0))
 \\
=&  \OmS'(N_1,N_2,N_3, \text{min}(c N_3, dN_1) - N_4)
= \OmS'(N_1,N_2,N_3, \text{min}(2 N_3, N_1) - N_4)\, .
\end{split}
\ee
This gives in particular
\ben \label{e4node1}
&& \OmS(1,1,1,1)=\OmS'(1,1,1,0), \quad \OmS(1,1,1,0)=\OmS'(1,1,1,1)\nonumber \\
&& \OmS(1,1,1,N) = 0 = \OmS'(1,1,1,N) \quad \text{for} \quad N\ge 2\, .
\een

We now proceed to verify \eqref{ecomp}. One finds using \eqref{essp1}: 
\begin{eqnarray}
\label{eq:invQ4}
Q(1,1,1,0) &=&  1/y^8 + 2/y^6 + 3/y^4 + 4/y^2 +5+ 4 y^2 + 3 y^4 + 2 y^6 + y^8\nonumber \\
Q(1,1,1,1)&=&y^{-8}+3y^{-6}+5y^{-4}+7y^{-2}+9+7y^{2}+5y^{4}+3y^{6}+y^{8}
+\OmS(1,1,1,1) \non\\
Q(1,1,1,2)&=&y^{-6}+2y^{-4}+3y^{-2}+4+3y^{2}+2y^{4}+y^{6}  
+ \OmS(1,1,1,1) + \OmS(1,1,1,2) \non \\
Q(1,1,1,3) &=& \OmS(1,1,1,2) + \OmS(1,1,1,3) \non \\
Q(1,1,1,4) &=& \OmS(1,1,1,3) +\OmS(1,1,1,4)\, ,
\end{eqnarray}
and
\begin{eqnarray}
\label{eq:invQ4m}
Q'(1,1,1,2) &=& 1/y^8 + 2/y^6 + 3/y^4 + 4/y^2 + 5+ 4 y^2 + 3 y^4 + 2 y^6 + y^8
\nonumber \\
&& + \OmS'(1,1,1,1) + \OmS'(1,1,1,2) \nonumber \\
Q'(1,1,1,1)&=&y^{-8}+3y^{-6}+5y^{-4}+7y^{-2}+9+7y^{2}+5y^{4}+3y^{6}+y^{8}\non \\
&&+\OmS'(1,1,1,0)+\OmS'(1,1,1,1)  \\
Q'(1,1,1,0)&=&y^{-6}+2y^{-4}+3y^{-2}+4+3y^{2}+2y^{4}+y^{6}+ \OmS'(1,1,1,0)\, .\non 
\end{eqnarray}
Compatibility of these expressions with
\eqref{eq:invQ4}, \eqref{ecomp} follows directly from \eqref{e4node1} and \eqref{e4node2}.
In particular the last two equations of
\eqref{eq:invQ4} are consistent with \eqref{ecomp}, \eqref{e4node1}. 
We can also test the vanishing of $Q'(1,1,1,N)$ for $N\ge 3$. For $\zeta'$ given by
\eqref{ezetamut} with $(N_1,N_2,N_3,N_4)=(1,1,1,2-N)$, we get
\ben
Q'(1,1,1,3) &=& \OmS'(1,1,1,2)+\OmS'(1,1,1,3)\, , \nonumber \\
Q'(1,1,1,4) &=& \OmS'(1,1,1,3)+ \OmS'(1,1,1,4)\ .
\een
These vanish using \eqref{e4node1}.

Note that in the above analysis we have not explicitly used 
the values of $\OmS$ and $\OmS'$ or tested \eqref{e4node1}.
{}From direct analysis of 3-node and 4-node cyclic
quiver given in \cite{Bena:2012hf,Lee:2012sc,Lee:2012naa,Manschot:2012rx} 
we know that $\OmS(1,1,1,0;t)=0$
(as there is no loop) and $\OmS(1,1,1,1;t)=4$. 
Thus we have
\be \label{e4node2}
\OmS'(1,1,1,1;t) = \OmS(1,1,1,0;t)=0, \quad \OmS'(1,1,1,0;t) = \OmS(1,1,1,1;t)=4\, .
\ee
The value of $\OmS'(1,1,1,0;t)$ given in 
\cite{Bena:2012hf,Lee:2012sc,Lee:2012naa,Manschot:2012rx}
agrees with the result given 
above.
Vanishing of $\OmS'(1,1,1,1;t)$ can be
seen by direct analysis of the Higgs branch moduli space of this quiver.

\section{Examples of generalized quiver mutations} \label{sgen}

In this section we 
test the conjectured invariance of the Coulomb branch formula for generalized
quivers where the condition \eqref{OmSquivers} is relaxed.

\bigskip

\noindent {\bf Example 1}: 

\medskip

We consider the generalized Kronecker quiver with  $m\equiv \gamma_{12}>0$ arrows
 from node 1 to node 2, 
 with  $\OmS(k\gamma_1;y;t)$ 
 and $\OmS(\ell \gamma_2;y;t)$  
 given by  arbitrary symmetric Laurent polynomials and 
$\OmS(\gamma)=0$ otherwise.   
In
the chamber $\zeta_1<0<\zeta_2$ the total index for charge $\gamma$ coincides
with $\OmS(\gamma)$ as there are no bound states with two or more centers. 
The index
in the other chamber  $\zeta_1>0>\zeta_2$, which we shall denote by
$Q(N_1, N_2)$,  can be obtained using the wall-crossing formula. 
We shall define, as in \eqref{essp1}, 
\ben
\bOmS(\gamma;y; t) &=& \sum_{m|\gamma} {1\over m} {y-y^{-1}\over y^m - y^{-m}} \, 
\OmS(\gamma/m; y^m; t^m)\, , \nonumber \\
\bQC(\gamma;y; t) &=& \sum_{m|\gamma} {1\over m} {y-y^{-1}\over y^m - y^{-m}} \, 
\QC(\gamma/m; y^m; t^m)\, ,
\een
and drop the arguments $y$ and $t$ from  $\bOmS$ to avoid cluttering. 
Using the shorthand notation $Q(p,q)$ for
$\QC(p\gamma_1+q\gamma_2;\zeta;y;t)$ etc.
the wall-crossing formula then takes the form
\be \label{ewallc}
\prod_{p,q\atop p/q \downarrow} \exp\left[ \bar Q(p,q) e_{p,q}\right]
= \exp\left[\sum_\ell \bOmS(\ell\gamma_2) e_{0,\ell}\right]
\exp\left[\sum_k \bOmS(k\gamma_1) e_{k,0}\right] \, 
\, ,
\ee
where $e_{p,q}$ are elements of an algebra satisfying the commutation relation
\ben
&& \left[ e_{p,q}, e_{p',q'}\right] = \kappa(\gamma, \gamma') \, e_{p+p',q+q'}, \quad
\gamma\equiv p\gamma_1+q\gamma_2, \quad \gamma'\equiv p'\gamma_1+q'\gamma_2, \nonumber\\
&&
\kappa(\gamma, \gamma')\equiv {(-y)^{\langle\gamma, \gamma'\rangle} - 
(-y)^{-\langle\gamma, \gamma'\rangle}\over y - y^{-1}}\, .
\een
The product over $p,q$  
runs over non-negative integers $p,q$ and
symbol $p/q\downarrow$ on the left hand side of \eqref{ewallc} implies that the product is
ordered such that the ratio $p/q$ decreases from left to right. 
If $p/q=p'/q'$ then the order is irrelevant since
$e_{p,q}$ and $e_{p',q'}$ will commute. Taking the $p=0$ terms on the left hand side
to the right hand side and using the fact that $\bar Q(0,\ell)=\bOmS(\ell\gamma_2)$,
we can express \eqref{ewallc} as
\be \label{ewallcmod} 
\begin{split}
 \prod_{p,q\atop p\ne 0, \, p/q \downarrow}\!\!\!\! \exp\left[ \bar Q(p,q) e_{p,q}\right]
=& \exp\left[\sum_\ell \bOmS(\ell\gamma_2) e_{0,\ell}\right] \,
 \exp\left[\sum_k \bOmS(k\gamma_1) e_{k,0}\right] \, 
\exp\left[-\sum_\ell \bOmS(\ell\gamma_2) e_{0,\ell}\right] \, .
\end{split}
\ee

Under generalized mutation with respect to the node 2, we have
$\gamma'_{12}=-\gamma_{12}$ and $\zeta'_1<0<\zeta'_2$. The effect of reversal of the sign
of $\zeta_i$'s will be to change the order of the products on both sides of
\eqref{ewallcmod}. On the other hand the effect of changing the 
sign of $\gamma_{12}$ is 
that the corresponding generators $e'_{p,q}$ which replace $e_{p,q}$ in \eqref{ewallcmod}
will satisfy a commutation relation similar to
that of $e_{p,q}$ but with an extra minus sign on the right hand side. 
This means that $-e'_{p,q}$'s will satisfy the same commutation relations as $e_{p,q}$'s.
Thus we can write an equation similar to that of \eqref{ewallcmod} with the
order of products reversed on both sides, $\bar Q(p,q)$ replaced by $\bar Q'(p,q)$ and
$e_{p,q}$ replaced by $-e_{p,q}$:
\be \label{ewallcpmod}
\begin{split}
 \prod_{p,q\atop p\ne 0, \, p/q \uparrow} &\exp\left[ -\bar Q'(p,q) e_{p,q}\right]
= \\&
 \exp\left[\sum_\ell \bOmS(\ell\gamma_2) e_{0,\ell}\right] \, \exp\left[-\sum_k
\bOmS(k\gamma_1) e_{k,0
}\right] \,  \exp\left[-\sum_\ell \bOmS(\ell\gamma_2) e_{0,\ell}\right] 
\, .
\end{split}
\ee
Taking the inverse of this has the effect of reversing the order of the products and 
changing  the
signs of $e_\gamma$'s in the exponent. The resulting equation is identical to that of 
\eqref{ewallcmod} 
 with $\bar Q(p,q)$ replaced by $\bar Q'(p,q)$,
showing that $\bar Q(p,q)=\bar Q'(p,q)$\cite{Manschot:2010qz}. 
Mutation invariance however
requires us to prove a different equality, namely 
$\bar Q'(p,q) = \bar Q(p, M\gamma_{12}p-q)$ where
\be
M \equiv \sum_\ell \ell^2 \OmS(\ell\gamma_2; y=1; t=1)\, .
\ee

To proceed, we shall assume that as a consequence of \eqref{ewallcmod} we have
\be \label{evanish}
Q(p,q) = 0 \quad \text{for} \quad q> M\gamma_{12}p\, .  
\ee
Later we shall prove this relation. Assuming this to be true, we define
$p'=p, \quad q'=M\gamma_{12}p-q$ (or equivalently $p=p'$, $q= M\gamma_{12}p'-q'$)
which are both non-negative for $p\ge 0$, 
$0\le q\le M\gamma_{12}p$ and note that $p'/q'$ are ordered in increasing order if $p/q$
are ordered in the decreasing order. Then we can express \eqref{ewallcmod} as
\ben \label{ewallcnew}
&& \prod_{p',q'\atop p'\ne 0, \, p'/q' \uparrow} \exp\left[ \bar Q(p',M\gamma_{12}p'-q') e_{p',M\gamma_{12}p'-q'}\right]
\nonumber \\ &=& \exp\left[\sum_\ell 
\bOmS(\ell\gamma_2) e_{0,\ell}\right] \,
\exp\left[\sum_k \bOmS(k\gamma_1) e_{k,0}\right] \, \exp\left[-\sum_\ell 
\bOmS(\ell\gamma_2) e_{0,\ell}\right]
\, .
\een
Since $p',q'$ are dummy indices we can change them to $p,q$ on the
left hand side. 
Furthermore notice that $e_{p,M\gamma_{12}p-q}$'s and 
$-e_{p,q}$'s have isomorphic
algebra for different $p,q$.  Thus we can replace $e_{p,M\gamma_{12}p-q}$
by $-e_{p,q}$ on both sides without changing the basic content of the
equations. This gives
\be
\begin{split} \label{ewallcnewp}
 \prod_{p,q\atop p\ne 0, \, p/q \uparrow}& \exp\left[ -\bar Q(p,M\gamma_{12}p-q) e_{p,q}\right] \\
=&  \exp\left[-\sum_\ell \bOmS(\ell\gamma_2) e_{0,-\ell}\right]\, 
\exp\left[-\sum_k \bOmS(k\gamma_1) e_{k,M\gamma_{12}k}\right] \, 
\exp\left[\sum_\ell \bOmS(\ell\gamma_2) e_{0,-\ell}\right]
\, .
\end{split}
\ee
Thus the proof of mutation
symmetry $\bar Q'(p,q)=\bar Q(p,M\gamma_{12}p-q)$ reduces to proving the equality of
the right hand sides of \eqref{ewallcpmod} and \eqref{ewallcnewp}. This is the task we shall
undertake now. For this we define 
\be \label{edefuv}
U \equiv \exp\left[\sum_\ell 
\bOmS(\ell\gamma_2) e_{0,\ell}\right] , \qquad 
V \equiv \exp\left[-\sum_\ell \bOmS(\ell\gamma_2) e_{0,-\ell}\right]\, ,
\ee
and express eqs.\eqref{ewallcpmod} and \eqref{ewallcnewp} as
\be \label{eonea}
\prod_{p,q\atop p\ne 0, \, p/q \uparrow} \exp\left[ -\bar Q'(p,q) e_{p,q}\right]
= \prod_k \exp\left[-
\bOmS(k\gamma_1) U \, e_{k,0
} U^{-1} \right]\, ,
\ee
and
\be \label{etwob}
\prod_{p,q\atop p\ne 0, \, p/q \uparrow} \exp\left[ -\bar Q(p,M\gamma_{12}p-q) e_{p,q}\right] 
= \prod_k \exp\left[- \bOmS(k\gamma_1) V \, e_{k,M\gamma_{12}k} V^{-1}\right] \, .
\ee
Note that the order of terms in the product over $k$ on the right hand sides of these two equations
is irrelevant since the terms for different $k$ commute.
Thus the equality of the right hand side of the two expressions require us to prove that
$U e_{k,0} U^{-1}=V e_{k, M\gamma_{12}k} V^{-1}$.

Now suppose we combine all the factors on either side of \eqref{eonea} and \eqref{etwob}
using the Baker-Campbell-Hausdorff formula, and consider the coefficients of $e_{1,s}$
in the exponent. 
On the left hand sides of \eqref{eonea} and \eqref{etwob},  these are determined in terms of $\bar Q'(1,q)$ and $\bar Q(1, M\gamma_{12}-q)$
respectively. Since we have already proved the equality of $\bar Q'(1,q)$ and
$\bar Q(1,M\gamma_{12}-q)$ with the help of semi-primitive wall-crossing formula, we see that 
the coefficients of $e_{1,s}$ in the exponent on the left hand sides are equal. On the
other hand since $U e_{k,0} U^{-1}$ and $V e_{k, M\gamma_{12}k} V^{-1}$ are linear combinations
of $e_{k, q}$, on the right hand sides the coefficient of $e_{1,s}$ in the exponents
are given by the terms proportional to $U e_{1,0}U^{-1}$ and 
$V e_{1,M\gamma_{12}} V^{-1}$, respectively. 
Thus the equality of the coefficients of $e_{1,s}$ in the exponent of
the two left hand sides imply that
\be 
U e_{1,0} U^{-1} = V e_{1,M\gamma_{12}} V^{-1}\, .
\ee
Now note that if we had considered a  
Kronecker quiver with nodes carrying charges 
$k\gamma_1$ for fixed $k$ and and $\ell\gamma_2$ for different $\ell>0$,
the semi-primitive wall-crossing formula would have given
the equality of this with a quiver whose nodes carry charges 
$k\gamma_1 + M\gamma_{12}k\gamma_2$
and $-\ell\gamma_2$ for dimension vector $(1,N)$. On the other hand such a quiver is equivalent 
to the one we are considering with $\OmS(r\gamma_1)=0$ for $r\ne k$, and we can use
\eqref{eonea}, \eqref{etwob} for such a quiver. 
In this case 
$\bar Q(p,q)$ and $\bar Q'(p, M\gamma_{12}p-q)$ would vanish for $1\le p\le k-1$ and for $p=k$ they would
be equal due to the generalized mutation invariance of the rank $(1,N)$ quiver. On the
right hand sides of the corresponding eqs.\eqref{eonea} and \eqref{etwob} the $e_{k,q}$ in
the exponent come from the $U e_{k,0} U^{-1}$ and $V e_{k,M\gamma_{12}k} V^{-1}$
terms, with $U$ and $V$ given by the same expressions \eqref{edefuv} as the original
quivers. Thus we conclude that
\be 
U e_{k,0} U^{-1} = V e_{k,M\gamma_{12}k} V^{-1}\, .
\ee
Since this is valid for every $k$, we see that the right hand sides of \eqref{eonea} and
\eqref{etwob} are equal for the original quiver. 
This in turn proves the equality of the left hand sides and hence the
desired relation
\be
\bar Q(p,M\gamma_{12}p-q) = \bar Q'(p,q)\, .
\ee

Finally,  we prove \eqref{evanish} as follows. 
From the
analysis of the rank $(1,N)$ case we know that $\bar Q'(1,q)$ vanishes for $q>M\gamma_{12}$.
With the help of \eqref{eonea} we can translate this to a statement that
$U e_{1,0} U^{-1}$ is a linear combination of $e_{1,q}$ for $0\le q\le M\gamma_{12}$. Generalizing this
to the quiver whose nodes carry charges $k\gamma_1$ and $\gamma_2$ we can
conclude that $U e_{k,0} U^{-1}$ is a linear combination of $e_{k,q}$ for $0\le q\le M\gamma_{12}k$.
Eq.\eqref{eonea} then shows that $\bar Q'(p,q)$ 
vanishes for $q>M\gamma_{12}p$. Equality of $Q(p,q)$ and $Q'(p,q)$, discussed below
\eqref{ewallcpmod} independent of the validity of generalized mutation symmetry, 
then leads to \eqref{evanish}.

We shall now test this for some specific choices of single-centered indices, namely
\be
\OmS(\gamma_1)=p_1, \quad 
\OmS(\gamma_2)=q_1,\quad \OmS(2\gamma_2)=q_2\ , \quad p_1,q_1,q_2\ge 0\, ,
\ee
with all other single-centered indices vanishing.   Generalized mutation  invariance
with respect to the node 2  requires that the
generating function 
\be
\FF(N_1,q; y;t) = \sum_{N_2\geq 0} \QC(N_1 \gamma_1 + N_2\gamma_2; \zeta; y;t)\, q^{N_2}
\ee
 satisfies the functional equation
\be
\label{funsemiprim}
q^{m N_1 M} \FF(N_1,1/q;y;t) = \FF(N_1,q;y;t)\ .
\ee
where $M\equiv q_1+4 q_2>0$. This equation holds for $N_1=1$
by assumption.  
Using the generalized semi-primitive formulae established in \cite{Manschot:2010qz}, we can test this property for $N_1=2$
or $N_1=3$. For simplicity we restrict to $N_2=2$, $1\leq m\leq 3$ and set $y=t=1$. We have computed 
$\FF(2,q)$ for the values of $(m,p_1,p_2,q_1,q_2)$ displayed in table \ref{fig_kron}, and found
that \eqref{funsemiprim} was indeed obeyed.

\begin{table}
$$
\begin{array}{|c|c|r|r|}\hline
m &p_1,q_1,q_2 & \FF(1,q) & \FF(2,q)\\ \hline
1 & 1,1,0 & 1+q & 0\\
1 & 1,2,0 & (1+q)^2 & 0\\
1 & 1,3,0 & (1+q)^3 & q^3 \\
1 & 2,1,0 & 2(1+q) & q \\
1 & 2,2,0 & 2(1+q)^2 & 2q(1-q+q^2)\\
1 & 2,3,0 & 2(1+q)^3 & q(3-6q+14q^2-6q^3+3q^4)\\
1 & 3,1,0 & 3(1+q) & 3q\\
1 & 3,2,0 & 3(1+q)^2 & 6q(1-q+q^2)\\
1 & 3,3,0 & 3(1+q)^3 & 3q(3-6q+13q^2-6q^3+3q^4)\\
\hline
2 & 1,1,0 & (1-q)^2 & q(1+q^2)\\
2 & 1,2,0 & (1-q)^4 & q(2-4q+22q^2-20q^3+22q^4-4q^5+2q^6)\\
2 & 2,1,0 & 2(1-q)^2 & 2q(3-2q+3q^2)\\
2 & 2,2,0 & 2(1-q)^4 & 4q(3-8q+29q^2-28q^3+29q^4-8q^5+3q^6)\\
\hline 
3 & 1,1,0 & (1+q)^3 & q(3-6q+13q^2-6q^3+3q^4)\\
3 & 1,2,0 & (1+q)^6 & 2q(3-15q+85 q^2-165 q^3+351 q^4-337 q^5+351 q^6 + \dots + 3 q^{10})\\ \hline
\end{array}
$$
\caption{Generating functions of $\QC(\gamma_1+N\gamma_2)$ and $\QC(2\gamma_1+N\gamma_2)$ for the generalized Kronecker quiver with $\OmS(\gamma_1)=p_1,\OmS(\gamma_2)=q_1,\OmS(2\gamma_2)=q_2$. The symmetry under $q\to 1/q$ shows mutation invariance in these cases.
\label{fig_kron}}
\end{table}

In this case, we can also test whether the conditions \eqref{eposcon0} can be relaxed.
Let us set $p_2=q_2=0$, $m=1$ for simplicity, and try 
$q_1=-1$. .
The semi-primitive partition function
\be
\FF(1,q)=\frac{p_1}{1+q}
\ee
is multiplied by $q$ 
under $q\to 1/q$  
but its rank 2 counterpart, computed using the formulae in \cite{Manschot:2010qz}, is not
multiplied by $q^2$ under $q\to 1/q$: 
\be
\FF(2,q)=\frac{p_1 q (1-p_1-(p_1+1)q^2)}{2(1-q)^2(1-q^4)}\ .
\ee
This illustrates the importance of the assumption that the mutating node 
must carry positive $\OmS$.

\bigskip

\noindent{\bf Example 2:} We consider a three node quiver of rank $(N_1, N_2, N_3)$ 
with 
$\gamma_{12}=\gamma_{32}=a=1$ and 
$\gamma_{31}=c=2$, and take the invariants
$\OmS(\ell\gamma_1)$ and $\OmS(\ell\gamma_3)$ to be generic functions of $\ell$, $y$ and $t$
and $\OmS(\ell\gamma_2;y;t)$ for different integers $\ell$
to be specific functions of $y$ and $t$ to be described below.
All other $\OmS(\gamma;y;t)$'s will be taken to vanish. 
For the FI parameters, we take
\be \label{efayet}
\zeta_1=(3\, N_2+.1)/N_1, \quad \zeta_2=-8, \quad \zeta_3=(5\, N_2-.1)/N_3\, .
\ee
Under mutation with respect to the node 2, we get
\be
\gamma_1' = \gamma_1 + M\, \gamma_2, \quad 
 \gamma_2'=-\gamma_2, \quad
\gamma_3'=\gamma_3 + M\, \gamma_2, \quad
\ee
\be 
\gamma'_{12}=-a\ ,\quad \gamma'_{23}=a\ ,\quad \gamma_{31}=c
\ee
where
\be
M = \sum_{\ell\ge 1} \ell^2 \OmS(\ell \gamma_2; y=1; t=1)\, .
\ee
Then 
\be 
N_1 \gamma_1 + N_2 \gamma_2 + N_3\gamma_3=
N_1 \gamma_1' + (M N_1+ M N_3- N_2) \gamma_2' + N_3 \gamma'_3\, .
\ee
The $\OmS'$'s for the mutated quiver are given by
\be \label{eommut}
\OmS'(\ell\gamma_1';y;t) = \OmS(\ell\gamma_1; y; t), \quad \OmS'(\ell\gamma_3';y;t) 
= \OmS(\ell\gamma_3; y; t), \quad \OmS'(\ell\gamma_2';y;t) = \OmS(\ell\gamma_2; y; t)\, .
\ee
Finally the FI parameters of the mutated quiver are
\be
\zeta_1' = (3\, N_2+.1)/N_1 - 8 M, \quad \zeta_3' = (5\, N_2-.1)/N_3- 8 M , \quad \zeta_2'
=8\, .
\ee
As before we 
denote $\QC(N_1\gamma_1+N_2\gamma_2+N_3\gamma_3;\zeta;y;t)$ by
$Q(N_1, N_2, N_3)$ and similarly for the mutated quiver. 
Also $\OmS(\gamma)$ without any other argument will denote $\OmS(\gamma;y;t)$.
The expected relationship
between $Q$ and $Q'$ then takes the form:
\be \label{eqqp}
Q(N_1, N_2, N_3) = Q'(N_1, M N_1 + M N_3 - N_2, N_3)\, .
\ee

We shall now consider several choices for the single-centered indices $\OmS(\ell\gamma_2; y; t)$. 

\medskip

\noindent{\bf (a):} $\OmS(\gamma_2; y; t) = 2, \quad \OmS(\ell\gamma_2; y; t) = 0$ for $\ell>1$.
In this case $M=2$, and the relation \eqref{eqqp} takes the form
\be \label{ereqd}
Q(N_1, N_2, N_3) = Q'(N_1, 2 N_1 + 2 N_3 - N_2, N_3)\, .
\ee
Explicit calculation gives
\ben
&& Q(1,2,1) = -(y^{-1}+ y) (y^{-2}+4 + y^2) \, \OmS(\gamma_1)\OmS(\gamma_3), \non\\ &&
Q'(1,2,1) = -(y^{-1}+ y) (y^{-2}+4 + y^2) \, \OmS'(\gamma_1')\OmS'(\gamma_3'), \non\\
&& Q(1,3,1) = 2 (y^{-2}+1 + y^2) \, \OmS(\gamma_1)\OmS(\gamma_3), \quad
Q'(1,1,1) = 2 (y^{-2}+1 + y^2) \, \OmS'(\gamma_1')\OmS'(\gamma_3'), \non\\
&& Q(1,4,1) = - (y^{-1}+ y) \, \OmS(\gamma_1)\OmS(\gamma_3), \quad
Q'(1,0,1) = - (y^{-1}+ y) \, \OmS'(\gamma_1')\OmS'(\gamma_3'), \non\\
&& Q(1,3,2) = (y^{-6} + y^{-4} + y^{-2} + 1 + y^2 + y^4+ 
     y^6) \non \\
     && \qquad \qquad \times 
     \bigg\{\OmS(\gamma_3; y; t)^2 - \OmS(\gamma_3; y^2; t^2) -
     2 (y^{-1} + y)\,  \OmS(2\gamma_3;  y; t)\bigg\} \, \OmS(\gamma_1; y; t)
\non\\
&& Q'(1,3,2) = (y^{-6} + y^{-4} + y^{-2} + 1 + y^2 + y^4+ 
     y^6) \non \\
     && \qquad \qquad \times 
     \bigg\{\OmS'(\gamma_3'; y; t)^2 - \OmS'(\gamma_3'; y^2; t^2) -
     2 (y^{-1} + y)\,  \OmS'(2\gamma_3';  y; t)\bigg\} \, \OmS'(\gamma_1'; y; t)\, .
\een
These results are in agreement with the generalized mutation hypothesis  \eqref{ereqd}.

\medskip

\noindent{\bf (b):} $\OmS(\gamma_2; y; t) = 3, \quad \OmS(\ell\gamma_2; y; t) = 0$ for $\ell>1$.
In this case $M=3$, and the relation \eqref{eqqp} takes the form
\be \label{ereqda}
Q(N_1, N_2, N_3) = Q'(N_1, 3 N_1 + 3 N_3 - N_2, N_3)\, .
\ee
Explicit calculation gives
\ben
&& Q(1,2,1) = -3 \, (y^{-3} + 4 y^{-1}+4 y + y^3) \, \OmS(\gamma_1)\OmS(\gamma_3), \non\\ &&
Q'(1,4,1) = -3 \, (y^{-3} + 4 y^{-1}+4 y + y^3) \, \OmS'(\gamma_1')\OmS'(\gamma_3'), \non\\
&& Q(1,3,1) = (y^{-4} + 10 y^{-2} + 10  + 10 y^2 + y^4) \, \OmS(\gamma_1)\OmS(\gamma_3), \non\\
&&
Q'(1,3,1) = (y^{-4} + 10 y^{-2} + 10  + 10 y^2 + y^4) \, \OmS'(\gamma_1')\OmS'(\gamma_3'), \non\\
&& Q(1,4,1) = - 3 \, (y^{-3} + 4 y^{-1}+4 y + y^3) \, \OmS(\gamma_1)\OmS(\gamma_3), \non\\
&&
Q'(1,2,1) = - 3 \, (y^{-3} + 4 y^{-1}+4 y + y^3) \, \OmS'(\gamma_1')\OmS'(\gamma_3'), \non\\
\een
in agreement with the generalized mutation hypothesis  \eqref{ereqda}.

\medskip

\noindent{\bf (c):} $\OmS(\gamma_2; y; t) = y^2+1+ y^{-2}, \quad 
\OmS(\ell\gamma_2; y; t) = 0$ for $\ell>1$.
In this case $M=3$, and the relation \eqref{eqqp} takes the form
\be \label{ereqdb}
Q(N_1, N_2, N_3) = Q'(N_1, 3 N_1 + 3 N_3 - N_2, N_3)\, .
\ee
Explicit calculation gives
\ben
&& Q(1,2,1) = -(2 y^{-5}+ 5 y^{-3} + 8 y^{-1} + 8 y + 5 y^3 + 2 y^5) \, \OmS(\gamma_1)\OmS(\gamma_3), \non\\ &&
Q'(1,4,1) = -(2 y^{-5}+ 5 y^{-3} + 8 y^{-1} + 8 y + 5 y^3 + 2 y^5) \, \OmS'(\gamma_1')\OmS'(\gamma_3'), \non\\
&& Q(1,3,1) = (y+y^{-1})^4 (y^2 + y^{-2}) \, \OmS(\gamma_1)\OmS(\gamma_3), \non\\
&&
Q'(1,3,1) = (y+y^{-1})^4 (y^2 + y^{-2}) \, \OmS'(\gamma_1')\OmS'(\gamma_3'), \non\\
&& Q(1,4,1) = - (2 y^{-5}+ 5 y^{-3} + 8 y^{-1} + 8 y + 5 y^3 + 2 y^5) \, \OmS(\gamma_1)\OmS(\gamma_3), \non\\
&&
Q'(1,2,1) = - (2 y^{-5}+ 5 y^{-3} + 8 y^{-1} + 8 y + 5 y^3 + 2 y^5) \, \OmS'(\gamma_1')\OmS'(\gamma_3'), \een
in agreement with the generalized mutation hypothesis  \eqref{ereqdb}.

\medskip

\noindent{\bf (d):} $\OmS(\gamma_2; y; t) = 4, \quad \OmS(\ell\gamma_2; y; t) = 0$ for $\ell>1$.
In this case $M=4$, and the relation \eqref{eqqp} takes the form
\be \label{ereqdc}
Q(N_1, N_2, N_3) = Q'(N_1, 4 N_1 + 4 N_3 - N_2, N_3)\, .
\ee
Explicit calculation gives
\ben
&& Q(1,4,1) = -(y^{-5} + 17 y^{-3} + 53 y^{-1} + 53 y + 17 y^3 + y^5) 
\, \OmS(\gamma_1)\OmS(\gamma_3), \non\\
&&
Q'(1,4,1) = -(y^{-5} + 17 y^{-3} + 53 y^{-1} + 53 y + 17 y^3 + y^5)  
\, \OmS'(\gamma_1')\OmS'(\gamma_3')\, .
\een
These results are in agreement with the generalized mutation hypothesis  \eqref{ereqdc}.

\medskip

\noindent{\bf (e):} $\OmS(\gamma_2; y; t) = t+1/t, \quad \OmS(\ell\gamma_2; y; t) = 0$ for $\ell>1$.
In this case
$M=2$, and the relation \eqref{eqqp} takes the form
\be \label{ereqde}
Q(N_1, N_2, N_3) = Q'(N_1, 2 N_1 + 2 N_3 - N_2, N_3)\, .
\ee
Explicit calculation gives
\ben
&& Q(1,2,1) = -(y^{-1}+ y) (t^{-2}+t^2 +y^{-2}+2 + y^2) \, \OmS(\gamma_1)\OmS(\gamma_3), \non\\ &&
Q'(1,2,1) = -(y^{-1}+ y) (t^{-2}+t^2 +y^{-2}+2 + y^2) \, \OmS'(\gamma_1')\OmS'(\gamma_3'), \non\\
&& Q(1,3,1) = (t^{-1}+t) (y^{-2}+1 + y^2) \, \OmS(\gamma_1)\OmS(\gamma_3), \non\\ &&
Q'(1,1,1) = (t^{-1}+t) (y^{-2}+1 + y^2) \, \OmS'(\gamma_1')\OmS'(\gamma_3'), \non\\
&& Q(1,4,1) = - (y^{-1}+ y) \, \OmS(\gamma_1)\OmS(\gamma_3), \quad
Q'(1,0,1) = - (y^{-1}+ y) \, \OmS'(\gamma_1')\OmS'(\gamma_3'), \non\\
&& Q(1,3,2) = {1\over 2} (t + t^{-1}) (y^{-6} + y^{-4} + y^{-2} + 1 + y^2 + y^4+ 
     y^6) \non \\
     && \qquad \qquad \times 
     \bigg\{\OmS(\gamma_3; y; t)^2 - \OmS(\gamma_3; y^2; t^2) -
     2 (y^{-1} + y)\,  \OmS(2\gamma_3;  y; t)\bigg\} \, \OmS(\gamma_1; y; t)
\non\\
&& Q'(1,3,2) = {1\over 2} (t + t^{-1})  (y^{-6} + y^{-4} + y^{-2} + 1 + y^2 + y^4+ 
     y^6) \non \\
     && \qquad \qquad \times 
     \bigg\{\OmS'(\gamma_3'; y; t)^2 - \OmS'(\gamma_3'; y^2; t^2) -
     2 (y^{-1} + y)\,  \OmS'(2\gamma_3';  y; t)\bigg\} \, \OmS'(\gamma_1'; y; t)\, . \non \\
\een
These results are in agreement with the generalized mutation hypothesis  \eqref{ereqde}.

\medskip

\noindent{\bf (f):} $\OmS(\gamma_2; y; t) = 0, \quad \OmS(2\gamma_2; y; t) = 1, 
\quad \OmS(\ell\gamma_2; y; t) = 0$ for $\ell>2$.
In this case $M=4$, and the relation \eqref{eqqp} takes the form
\be \label{ereqdf}
Q(N_1, N_2, N_3) = Q'(N_1, 4 N_1 + 4 N_3 - N_2, N_3)\, .
\ee
Explicit calculation gives
\ben
&& Q(1,4,1) = -(y^{-5} + 2 y^{-3} + 4 y^{-1} + 4 y + 2 y^3 + y^5) 
\, \OmS(\gamma_1)\OmS(\gamma_3), \non\\
&&
Q'(1,4,1) = -(y^{-5} + 2 y^{-3} + 4 y^{-1} + 4 y + 2 y^3 + y^5)  
\, \OmS'(\gamma_1')\OmS'(\gamma_3')\, .
\een
These results are in agreement with the generalized mutation hypothesis \eqref{ereqdf}.

\medskip

\noindent{\bf (g):} We end this series of examples with a choice of
$\OmS$ which violates condition i) on page \pageref{condi}, but which
preserves the mutation symmetry at the level of numerical
DT-invariants. We mentioned earlier this possibility in Section \ref{sgenmut}. We take $\OmS(\gamma_2;y;t)=-1$ and
$\OmS(2\gamma_2;y;t)=1$. We may expect the generalized mutation to be a symmetry for $y=t=1$
since the generating functions $\FF(\vec N;q;\zeta;q;1;1)$ are symmetric
polynomials in $q$. In particular for this choice we have $M=3$ and
hence $Q(N_1, N_2, N_3)$ would have to be equal to $Q'(N_1,3 N_1+3N_3-N_2, N_3)$.
We find that while this does not hold for general $y$, it does hold for $y=t=1$. 
For example we have $Q(1,4,1)=Q'(1,2,1)=2$ at $y=1$. 

\medskip

\noindent{\bf Example 3:} Now we consider a three node quiver with
loop by choosing  $\gamma_{12}=2$, $\gamma_{23}=1$, and
$\gamma_{31}=5$.
We choose 
 $\OmS(\gamma_2;y;t)=2$, $\OmS(\ell\gamma_2;y;t)=0$ for $\ell>1$, and
leave $\OmS(N_1\gamma_1+N_2\gamma_2+N_3\gamma_3; y; t)$ arbitrary
except for the constraints imposed due to the restrictions mentioned at the end
of \S\ref{sintro}. This in particular will require $\OmS$ to vanish when either
$N_1$ or $N_3$ vanishes with other $N_i$'s being given by positive integers.
The choice of FI parameters remain the same as
in \eqref{efayet}:
\be \label{efayetrep}
\zeta_1=(3\, N_2+.1)/N_1, \quad \zeta_2=-8, \quad \zeta_3=(5\, N_2-.1)/N_3\, .
\ee

Under mutation with respect to the node 2, we get
\be
\gamma_1' = \gamma_1 + 4 \gamma_2, \quad
\gamma_2'=-\gamma_2, \quad
 \gamma_3'=\gamma_3,
\ee
\be 
N_1 \gamma_1 + N_2 \gamma_2 + N_3\gamma_3=
N_1 \gamma_1' + (4 N_1 - N_2) \gamma_2' + N_3 \gamma_3'\, .
\ee
The $\OmS'$'s for the mutated quiver for charge vectors proportional
to the basis vectors continue to be given by \eqref{eommut}.
For general charge vectors we get from \eqref{egench00}
\be \label{eomreln}
\OmS(N_1\gamma_1+N_2\gamma_2+N_3\gamma_3; y; t)
=\begin{cases} \OmS'(N_1\gamma_1'+(2 N_3-N_2)\gamma_2'+N_3\gamma_3'; y; t) \quad
\text{for} \quad 2 N_1 \ge N_3\\
\OmS'(N_1\gamma_1'+(4 N_1-N_2)\gamma_2'+N_3\gamma_3'; y; t) \quad
\text{for} \quad 2 N_1 < N_3
\end{cases}\, .
\ee
Finally the FI parameters of the mutated quiver are
\be
\zeta_1' = (3\, N_2+.1)/N_1 - 32  , \quad \zeta_3' = (5\, N_2-.1)/N_3, \quad \zeta_2'
=8 \, .
\ee
The mutated quiver has
\be
\gamma'_{12} = -2, \quad \gamma'_{13} = -1, \quad \gamma'_{23} = -1\, .
 \ee
and the expected relation is
\be \label{expect}
Q(N_1, N_2, N_3) = Q'(N_1, 4 N_1-N_2, N_3)\, .
\ee

Explicit calculation gives
\begin{eqnarray}
&& Q(1,2,1) = (y^{-4} + 5 y^{-2} + 6 + 5 y^2 + y^4) \, \OmS(\gamma_1;y;t)\OmS(\gamma_3;y;t) \non\\
&& \quad + \, \OmS(\gamma_1+\gamma_3;y;t)  + 2  \, \OmS(\gamma_1+\gamma_2+\gamma_3;y;t) 
 + \, \OmS(\gamma_1+2\gamma_2+\gamma_3;y;t) \non \\ 
&&Q'(1,2,1) = (y^{-4} + 5 y^{-2} + 6 + 5 y^2 + y^4)  \, \OmS'(\gamma_1';y;t)\OmS'(\gamma_3';y;t) \non\\
&& \quad +\, \OmS'(\gamma'_1+\gamma'_3;y;t)  + 2 \, \OmS'(\gamma'_1+\gamma'_2+\gamma'_3;y;t) 
 +\, \OmS'(\gamma'_1+2\gamma'_2+\gamma'_3;y;t) \non \\ 
 && Q(1,3,1) = 2\, (y^{-1} + y)^2 \, \OmS(\gamma_1;y;t)\OmS(\gamma_3;y;t) \non\\
&& \quad +\OmS(\gamma_1+\gamma_2+\gamma_3;y;t)  + 2  \, \OmS(\gamma_1+2\gamma_2+\gamma_3;y;t) 
 + \, \OmS(\gamma_1+3\gamma_2+\gamma_3;y;t) \non \\ 
&&Q'(1,1,1) = 2\, (y^{-1} + y)^2\, \OmS'(\gamma_1';y;t)\OmS'(\gamma_3';y;t)\, \non\\
&& \quad  + 2 \, \OmS'(\gamma'_1+\gamma'_3;y;t) 
 +\, \OmS'(\gamma'_1+\gamma'_2+\gamma'_3;y;t) \non \\
&& Q(1,2,2) =  {1\over 2}(y^{-2} + 1 + y^2) (y^{-2} + 4  + 
     y^2) \bigg\{(y^{-2} + 1 + y^2)\, \OmS(\gamma_3; y; 
       t)^2 \non \\ && - (y^{-2} - 1 + y^2)  \, \OmS(\gamma_3; y^2; t^2) - 2\, (y^{-2} - 1 + y^2)
        (y^{-1} + y) \, \OmS(2\gamma_3; y; t))\bigg\} \, \OmS(\gamma_1;y;t) \non\\
&& + (y^{-2}+1+y^2) \, \OmS(\gamma_3;y;t) \Big\{  \OmS(\gamma_1+\gamma_3;y;t)  + 2  \, \OmS(\gamma_1+\gamma_2+\gamma_3;y;t) \non \\&&
 + \, \OmS(\gamma_1+2\gamma_2+\gamma_3;y;t)\Big\} 
 + \, \OmS(\gamma_1+2\gamma_2+2\gamma_3;y;t) \non \\
&&Q'(1,2,2) = {1\over 2}(y^{-2} + 1 + y^2) (y^{-2} + 4  + 
     y^2) \bigg\{(y^{-2} + 1 + y^2)\, \OmS'(\gamma_3'; y; 
       t)^2 \non \\ && - (y^{-2} - 1 + y^2)  \, \OmS'(\gamma_3'; y^2; t^2) - 2\, (y^{-2} - 1 + y^2)
        (y^{-1} + y) \, \OmS'(2\gamma_3'; y; t))\bigg\} \, \OmS'(\gamma_1';y;t) \non\\
&& + (y^{-2}+1+y^2) \, \OmS'(\gamma_3';y;t) 
 \Big\{  \OmS'(\gamma'_1+\gamma'_3;y;t)  + 
 2  \, \OmS'(\gamma'_1+\gamma'_2+\gamma'_3;y;t) \non \\&&
 + \, \OmS'(\gamma'_1+2\gamma'_2+\gamma'_3;y;t)\Big\} 
 + \, \OmS'(\gamma'_1+2\gamma'_2+2\gamma'_3;y;t) \non 
\end{eqnarray}
 \begin{eqnarray}
 && Q(1,2,3) ={1\over 6} (y^{-2} + 4  + 
   y^2) \bigg[(y^{-2} + 1 + y^2)^3 \, \OmS(\gamma_3; y; t)^3 \non\\ &&
    - 
   3 (y^{-2} - 1 + y^2) (y^{-2} + 1 + y^2)^2 \, \OmS(\gamma_3; y; 
     t) \Big\{ \OmS(\gamma_3; y^2; t^2) + 
      2 (y^{-1} + y)\,  \OmS(2\gamma_3; y;t)\Big\} \non\\ && + 
   2 (y^{-6} +1 + y^6) \Big\{\OmS(\gamma_3; y^3; t^3) + 
      3 (y^{-2} + 1 + y^2) \, \OmS(3\gamma_3; y; t)\Big\}\bigg] \, \OmS(\gamma_1; y; t)\non\\ &&
+ {1\over 2} \bigg\{(y^{-2} + 1 + y^2)^2 \OmS(\gamma_3; y; 
     t)^2  - 
   (y^{-4} + 1 + y^4)\OmS(\gamma_3; y^2; t^2)  \non \\ &&
   - 
  2 (y^{-5} + y^{-3} + y^{-1} + y  + y^3 + y^5)  \OmS(2\gamma_3; y; t)
   \bigg\}  \non \\ &&
   \times \bigg\{ \OmS(\gamma_1+ \gamma_3; y; 
     t)   + 2 \OmS(\gamma_1+\gamma_2+\gamma_3; y; 
     t) + \OmS(\gamma_1+2\gamma_2+\gamma_3; y; 
     t) 
     \bigg\} \non \\ &&   + 
    (y^{-2} + 1 + y^2) \OmS(\gamma_3; y; 
     t) \OmS(\gamma_1+2\gamma_2+2\gamma_3; y; 
     t)  \non \\ &&
      + \OmS(\gamma_1+2\gamma_2+3\gamma_3; y; 
     t)  \non
     \end{eqnarray}
 \begin{eqnarray}
&&
Q'(1,2,3) ={1\over 6} (y^{-2} + 4  + 
   y^2) \bigg[(y^{-2} + 1 + y^2)^3 \, \OmS'(\gamma'_3; y; t)^3 \non\\ &&
    - 
   3 (y^{-2} - 1 + y^2) (y^{-2} + 1 + y^2)^2 \, \OmS'(\gamma'_3; y; 
     t) \Big\{ \OmS'(\gamma'_3; y^2; t^2) + 
      2 (y^{-1} + y)\,  \OmS'(2\gamma'_3; y;t)\Big\} \non\\ && + 
   2 (y^{-6} +1 + y^6) \Big\{\OmS'(\gamma'_3; y^3; t^3) + 
      3 (y^{-2} + 1 + y^2) \, \OmS'(3\gamma'_3; y; t)\Big\}\bigg] \, \OmS'(\gamma'_1; y; t)
      \non\\ &&
+ {1\over 2} \bigg\{(y^{-2} + 1 + y^2)^2 \OmS'(\gamma'_3; y; 
     t)^2  - 
   (y^{-4} + 1 + y^4)\OmS'(\gamma'_3; y^2; t^2)  \non \\ &&
   - 
  2 (y^{-5} + y^{-3} + y^{-1} + y  + y^3 + y^5)  \OmS'(2\gamma'_3; y; t)
   \bigg\}  \non \\ &&
   \times \bigg\{ \OmS'(\gamma'_1+ \gamma'_3; y; 
     t)   + 2 \OmS'(\gamma'_1+\gamma'_2+\gamma'_3; y; 
     t) + \OmS'(\gamma'_1+2\gamma'_2+\gamma'_3; y; 
     t) 
     \bigg\} \non \\ &&   + 
    (y^{-2} + 1 + yf^2) \OmS'(\gamma'_3; y; 
     t) \OmS'(\gamma'_1+2\gamma'_2+2\gamma'_3; y; 
     t)  \non \\ &&
      + \OmS'(\gamma'_1+2\gamma'_2+3\gamma'_3; y; 
     t) \, .
 \end{eqnarray}
Using \eqref{eomreln} we see that 
these results are in agreement with 
\eqref{expect}.

We have checked similar agreement for many other examples.

\section*{Acknowledgement}
We would like to thank Bernhard Keller, Gregory W. Moore,  Andy Neitzke
and Piljin Yi for 
inspiring discussions. Part of the reported results were obtained while J.M. was a postdoc of the Bethe Center for Theoretical 
Physics of Bonn University. This work was supported in part by the National Science Foundation under 
Grant No. PHYS-1066293 and the hospitality of the Aspen Center for Physics.
The work of A.S. was supported in part by DAE project 12-R\&D-HRI-5.02-0303 and the J.C. Bose fellowship of DST, Govt. of India.



\providecommand{\href}[2]{#2}\begingroup\raggedright\endgroup

\end{document}